\documentclass[12pt]{article}
\usepackage[colorlinks,linkcolor=Blue,citecolor=Blue,bookmarks,bookmarksnumbered]{hyperref}
\usepackage[scaled=0.85]{helvet}
\usepackage{amsmath,amssymb,accents,mathrsfs,XoohmE}
\usepackage{graphicx,color}
\usepackage{booktabs}
\usepackage{multirow}
\usepackage{placeins}

\def\rD{{\rm D}}
\def\rI{{\rm I}}
\def\rJ{{\rm J}}

\def\rL{{\rm L}}
\def\rR{{\rm R}}

\def\hi{{\hat\imath}}
\def\hj{{\hat\jmath}}
\def\hk{{\hat{k}}}
\def\hl{{\hat\ell}}

\def\vCent#1{\vcenter{\hbox{\hss#1\hss}}}
\def\be{\begin{equation}}
\def\ee{\end{equation}}
\newcommand{\bea}{\begin{eqnarray}}
\newcommand{\eea}{\end{eqnarray}}
\newcommand{\ena}{\end{eqnarray}}
\newcommand{\CN}{{\cal N}}

\def\bj#1{{}_{\rm #1}}                          

\def\pp{{\mathchoice
          {
              \kern 1pt%
              \raise 1pt
              \vbox{\hrule width5pt height0.4pt depth0pt
                    \kern -2pt
                    \hbox{\kern 2.3pt
                          \vrule width0.4pt height6pt depth0pt
                          }
                    \kern -2pt
                    \hrule width5pt height0.4pt depth0pt}%
                    \kern 1pt
           }
            {
              \kern 1pt%
              \raise 1pt
              \vbox{\hrule width4.3pt height0.4pt depth0pt
                    \kern -1.8pt
                    \hbox{\kern 1.95pt
                          \vrule width0.4pt height5.4pt depth0pt
                          }
                    \kern -1.8pt
                    \hrule width4.3pt height0.4pt depth0pt}%
                    \kern 1pt
            }
            {
              \kern 0.5pt%
              \raise 1pt
              \vbox{\hrule width4.0pt height0.3pt depth0pt
                    \kern -1.9pt  
                    \hbox{\kern 1.85pt
                          \vrule width0.3pt height5.7pt depth0pt
                          }
                    \kern -1.9pt
                    \hrule width4.0pt height0.3pt depth0pt}%
                    \kern 0.5pt
            }
            {
              \kern 0.5pt%
              \raise 1pt
              \vbox{\hrule width3.6pt height0.3pt depth0pt
                    \kern -1.5pt
                    \hbox{\kern 1.65pt
                          \vrule width0.3pt height4.5pt depth0pt
                          }
                    \kern -1.5pt
                    \hrule width3.6pt height0.3pt depth0pt}%
                    \kern 0.5pt
            }
        }}

\def\mm{{\mathchoice
                       {
                             \kern 1pt
               \raise 1pt    \vbox{\hrule width5pt height0.4pt depth0pt
                                  \kern 2pt
                                  \hrule width5pt height0.4pt depth0pt}
                             \kern 1pt}
                       {
                            \kern 1pt
               \raise 1pt \vbox{\hrule width4.3pt height0.4pt depth0pt
                                  \kern 1.8pt
                                  \hrule width4.3pt height0.4pt depth0pt}
                             \kern 1pt}
                       {
                            \kern 0.5pt
               \raise 1pt
                            \vbox{\hrule width4.0pt height0.3pt depth0pt
                                  \kern 1.9pt
                                  \hrule width4.0pt height0.3pt depth0pt}
                            \kern 1pt}
                       {
                           \kern 0.5pt
             \raise 1pt  \vbox{\hrule width3.6pt height0.3pt depth0pt
                                  \kern 1.5pt
                                  \hrule width3.6pt height0.3pt depth0pt}
                           \kern 0.5pt}
                       }}

\def\ad{{\kern0.5pt
                   \alpha \kern-5.05pt \raise5.8pt\hbox{$\textstyle.$}\kern
0.5pt}}

\def\bd{{\kern0.5pt
                   \beta \kern-5.05pt \raise5.8pt\hbox{$\textstyle.$}\kern
0.5pt}}

\def\qd{{\kern0.5pt
                   q \kern-5.05pt \raise5.8pt\hbox{$\textstyle.$}\kern
0.5pt}}
\def\Dot#1{{\kern0.5pt
     {#1} \kern-5.05pt \raise5.8pt\hbox{$\textstyle.$}\kern
0.5pt}}

\catcode`@=11
\def\un#1{\relax\ifmmode\@@underline#1\else
        $\@@underline{\hbox{#1}}$\relax\fi}
\catcode`@=12

\def\a{\alpha}
\def\b{\beta}
\def\c{\chi}
\def\d{\delta}

\def\f{\phi}

\def\i{\iota}
\def\j{\psi}
\def\k{\kappa}
\def\l{\lambda}
\def\m{\mu}
\def\n{\nu}
\def\o{\omega}

\def\r{\rho}
\def\s{\sigma}
\def\t{\tau}

\def\F{\Phi}

\def\J{\Psi}
\def\L{\Lambda}
\def\O{\Omega}
\def\P{\Pi}

\def\S{\Sigma}

\def\ca{{\cal A}}

\def\cf{{\cal F}}
\def\cg{{\cal G}}

\def\car{{\cal R}}

\def\dslash{\not{\hbox{\kern-2pt $\partial$}}}
\def\Dslash{\not{\hbox{\kern-4pt $D$}}}
\def\pslash{\not{\hbox{\kern-2.3pt $p$}}}
 \newtoks\slashfraction
 \slashfraction={.13}
 \def\slash#1{\setbox0\hbox{$ #1 $}
 \setbox0\hbox to \the\slashfraction\wd0{\hss \box0}/\box0 }
 
\def\kcr{{\hbox{\ro \char'170}}}                
\def\ktl{{\hbox{\ro \char'170}}}        
\def\ktr{{\hbox{\ro \char'170}}}        
\def\kbl{{\hbox{\ro \char'170}}}        
\def\kbr{{\hbox{\ro \char'170}}}        

\def\plpl{\raise-2pt\hbox{$\raise3pt\hbox{$_+$}\hskip-6.67pt\raise0.0pt
\hbox{$^+$}\hskip 0.01pt$}}
\def\mimi{\raise-2pt\hbox{$\raise3pt\hbox{$_-$}\hskip-6.67pt\raise0.0pt
\hbox{$^-$}\hskip 0.01pt$}} 

\def\bo{{\raise.15ex\hbox{\large$\Box$}}}               
\def\iff{\leftrightarrow}                               
\def\TH{{\raise.2ex\hbox{$\displaystyle \bigodot$}\mskip-4.7mu \llap H \;}}
\def\face{{\raise.2ex\hbox{$\displaystyle \bigodot$}\mskip-2.2mu \llap {$\ddot
        \smile$}}}                                      

\def\dt#1{\on{\hbox{\bf .}}{#1}}                
\def\Dot#1{\dt{#1}}

\def\Hat#1{\widehat{#1}}                        
\def\leftrightarrowfill{$\mathsurround=0pt \mathord\leftarrow \mkern-6mu
        \cleaders\hbox{$\mkern-2mu \mathord- \mkern-2mu$}\hfill
        \mkern-6mu \mathord\rightarrow$}
\def\dvec#1{\vbox{\ialign{##\crcr
        \leftrightarrowfill\crcr\noalign{\kern-1pt\nointerlineskip}
        $\hfil\displaystyle{#1}\hfil$\crcr}}}           
\def\dt#1{{\buildrel {\hbox{\LARGE .}} \over {#1}}}     

\def\sfrac#1#2{{\vphantom1\smash{\lower.5ex\hbox{\small$#1$}}\over
        \vphantom1\smash{\raise.4ex\hbox{\small$#2$}}}} 
\def\bfrac#1#2{{\vphantom1\smash{\lower.5ex\hbox{$#1$}}\over
        \vphantom1\smash{\raise.3ex\hbox{$#2$}}}}       
\def\afrac#1#2{{\vphantom1\smash{\lower.5ex\hbox{$#1$}}\over#2}}    

\def\opR{{\bm{\cal R}}}

\newcommand{\bm}[1]{{\boldsymbol{#1}}}

\def\ad{{\dot{\alpha}}}
\def\bd{{\dot{\beta}}}

 \font\rOpe=cmsy10                        
 \def\ktl{{\hbox{\rOpe\char'170}}}        
 \def\kbl{{\hbox{\rOpe\char'170}}}        
 \def\kcr{{\reflectbox{\rOpe\char'170}}}        
 \def\ktr{{\reflectbox{\rOpe\char'170}}}        
 \def\kbr{{\reflectbox{\rOpe\char'170}}}        
 \def\Border{\vbox{\hsize0pt
        \setlength{\unitlength}{1mm}
        \newcount\xco
        \newcount\yco
        \xco=-21
        \yco=12
        \begin{picture}(0,0)(-7.5,0)
        \put(\xco,\yco){$\ktl$}
        \advance\yco by-1
        {\loop
        \put(\xco,\yco){$\kcr$}
        \advance\yco by-2
        \ifnum\yco>-240
        \repeat
        \put(\xco,\yco){$\kbl$}}
        \xco=170
        \yco=12
        \put(\xco,\yco){$\ktr$}
        \advance\yco by-1
        {\loop
        \put(\xco,\yco){$\kcr$}
        \advance\yco by-2
        \ifnum\yco>-240
        \repeat
        \put(\xco,\yco){$\kbr$}}
        \put(-19.5,13){\scalebox{.6065}{%
         University of Maryland Center for String and Particle  Theory \&\ Physics Department%
        |University of Maryland Center for String and Particle  Theory \&\ Physics Department}}
        \put(-19.5,-241.5){\scalebox{.5835}{%
         Howard University Department of Physics and Astronomy%
        |Howard University Department of Physics and Astronomy%
        |Howard University Department of Physics and Astronomy}}
        \end{picture}
        \par\vskip-8mm}}
\definecolor{UMred}{rgb}{.9,.05,.2}
\definecolor{HUblue}{rgb}{.0,.3,.7}
 \def\UMbanner{\vbox{\hsize0pt
        \setlength{\unitlength}{.4mm}
        \thicklines\color{UMred}
        \begin{picture}(0,0)(-30,-10)
        \put(165,16){\line(1,0){4}}
        \put(170,16){\line(1,0){4}}
        \put(180,16){\line(1,0){4}}
        \put(175,0){\line(1,0){4}}
        \put(180,0){\line(1,0){4}}
        \put(185,0){\line(1,0){4}}
        \put(169,0){\line(0,1){16}}
        \put(170,0){\line(0,1){16}}
        \put(179,0){\line(0,1){16}}
        \put(180,0){\line(0,1){16}}
        \put(184,0){\line(0,1){16}}
        \put(185,0){\line(0,1){16}}
        \put(169,16){\oval(8,32)[bl]}
        \put(170,16){\oval(8,32)[br]}
        \put(179,0){\oval(8,32)[tl]}
        \put(185,0){\oval(8,32)[tr]}
        \color{HUblue}
        \put(167.75,-2){\line(1,0){4}}
        \put(172.75,-2){\line(1,0){4}}
        \put(177.75,-2){\line(1,0){4}}
        \put(182.75,-2){\line(1,0){4}}
        \put(167.75,-2){\line(0,-1){16}}
        \put(171.75,-2){\line(0,-1){16}}
        \put(172.75,-2){\line(0,-1){16}}
        \put(176.75,-2){\line(0,-1){16}}
        \put(181.75,-2){\line(0,-1){16}}
        \put(182.75,-2){\line(0,-1){16}}
        \put(181.75,-2){\oval(8,32)[bl]}
        \put(182.75,-2){\oval(8,32)[br]}
        \put(167.75,-18){\line(1,0){4}}
        \put(172.75,-18){\line(1,0){4}}
        \end{picture}
        \par\vskip-6.5mm
        \thicklines}}

\definecolor{Red}    {rgb}{0.90,0.00,0.12} 
\definecolor{Blue}   {rgb}{0.00,0.00,1.00} 
\definecolor{Green}  {rgb}{0.10,0.70,0.10} 
\definecolor{Turque} {rgb}{0.00,0.65,0.85} 
\definecolor{Orange} {rgb}{1.00,0.50,0.15} 
\definecolor{Magenta}{rgb}{1.00,0.00,1.00} 
\definecolor{Gold}   {rgb}{1.00,0.75,0.25} 
\definecolor{Seaweed}{rgb}{0.01,0.24,0.09} 
\definecolor{Purple} {rgb}{0.50,0.25,0.55} 
\definecolor{Brown}  {rgb}{0.43,0.26,0.32} 
\definecolor{grey1}  {rgb}{0.20,0.20,0.20} 
\definecolor{grey2}  {rgb}{0.40,0.40,0.40} 
\definecolor{grey3}  {rgb}{0.60,0.60,0.60} 
\definecolor{grey4}  {rgb}{0.80,0.80,0.80} 
\definecolor{grey5}  {rgb}{0.90,0.90,0.90} 
\def\C#1#2{{\ifcase#1\or
             \color{Red}\or \color{Green}\or \color{Blue}\or\
              \color{Turque}\or \color{Orange}\or \color{Magenta}\or 
               \color{Gold}\or \color{Seaweed}\or \color{Purple}\or
                \color{Brown}\or\color{grey1}\or\color{grey2}\or
                 \color{grey3}\else\color{grey4}\fi#2}}

\definecolor{Slate} {rgb}{0.00,0.45,0.55}

\def\Ft#1{\,\footnote{#1}}
\newdimen\parshift\parshift=\parindent
\catcode`@=11
 \long\def\@footnotetext#1{\insert\footins{\reset@font\footnotesize
           \interlinepenalty\interfootnotelinepenalty\splittopskip%
            \footnotesep\splitmaxdepth\dp\strutbox\floatingpenalty\@MM%
             \hsize\columnwidth\addtolength{\hsize}{-2\parindent}
              \@parboxrestore\protected@edef\@currentlabel%
              {\csname p@footnote\endcsname\@thefnmark}%
                \color@begingroup%
                 \@makefntext{\rule\z@\footnotesep\ignorespaces#1%
                  \@finalstrut\strutbox}%
                \color@endgroup}}
 \long\def\@makefntext#1{\hglue\parshift%
           \vbox{\noindent\baselineskip=11pt plus.5pt minus.5pt\hb@xt@0em{\hss\@makefnmark\kern1pt}#1}}
\catcode`@=12

\newskip\humongous \humongous=0pt plus 1000pt minus 1000pt
\def\caja{\mathsurround=0pt}

\newif\ifdtup
\def\panorama{\global\dtuptrue \openup2\jot \caja
        \everycr{\noalign{\ifdtup \global\dtupfalse
        \vskip-\lineskiplimit \vskip\normallineskiplimit
        \else \penalty\interdisplaylinepenalty \fi}}}

\makeatletter
\def\section{\@startsection{section}{1}{\z@}
        {3ex plus-1ex minus-.2ex}{1pt plus1pt}{\large\sf\bfseries\boldmath}}
\def\subsection{\@startsection{subsection}{2}{\z@}
         {1.5ex plus-1ex minus-.2ex}{0.01pt plus1pt}{\sf\slshape}}
\def\subsubsection{\@startsection{subsubsection}{3}{\z@}
          {1.5ex plus-1ex minus-.2ex}{0.01pt plus0.2pt}{\sf\boldmath}}
\def\paragraph{\@startsection{paragraph}{4}{\z@}
           {.75ex \@plus.5ex \@minus.2ex}{-2mm}{\sf\bfseries\boldmath}}
\makeatother

 \allowdisplaybreaks
 \seceq

\begin{document}

\thispagestyle{empty}
\vbox{\Border\UMbanner}
 \noindent{\small
 \today\hfill{PP-012-014 
 }}
  \vspace*{5mm}
 \begin{center}
{\LARGE\sf\bfseries
 Adinkra (In)Equivalence\\[-2mm]
 From Coxeter Group Representations:\\[2mm]
 A Case Study}
 \\[12mm]
{\large\sf\bfseries Isaac Chappell II$^*$\footnote{ichappel@umd.edu},~
                    S.\ James Gates, Jr.$^*$\footnote{gatess@wam.umd.edu},~
                     and\, 
                    T.\, H\"{u}bsch$^{\dag\ddag}$\footnote{thubsch@howard.edu}
                    }\\*[4mm]
\emph{
      \centering
  $^*$Center for String and Particle Theory, Dept.\ of Physics, \\[-2pt]
      University of Maryland, College Park, MD 20472,  
   \\[3pt]
      $^\dag$Dept.\ of Physics \&\ Astronomy, Howard University, Washington, DC 20059
   \\[1pt]
     $^\ddag$Dept.\ of Physics, University of Central Florida, Orlando, FL 
}
 \\*[30mm]
{\sf\bfseries ABSTRACT}\\[4mm]
\parbox{142mm}{\parindent=2pc\indent\baselineskip=14pt plus1pt
Using a \textsl{Mathematica$^{\sss\text{TM}}$} code, we present a straightforward
numerical ana\-lysis of the 384-dimensional solution space of signed permutation 
$4\times4$ matrices, which in sets of four provide representations of the ${\cal 
GR}(4,4)$ algebra, closely related to the ${\cal N}=1$ (simple) supersymmetry 
algebra in 4-dimensional spacetime. Following after ideas discussed in previous 
papers about automorphisms and classification of adinkras and corresponding 
supermultiplets, we make a new and alternative proposal to use equivalence 
classes of the (unsigned) permutation group $S_4$ to define distinct representations 
of higher dimensional spin bundles within the context of adinkras.  For this purpose, 
the definition of a dual operator akin to the well-known Hodge star is found to partition the space of these ${\cal GR}(4,4)$ representations into three suggestive classes.
 }
 \end{center}
\vfill
\vfill
\clearpage

\section{Introduction}
In a previous work \cite{Auto}, different ways to classify adinkras were considered.
These methods included Clifford algebras and certain binary linear block codes to
describe the marked topology types and isomorphisms between adinkras that faithfully
depict an unexpectedly vast collection: Refs.\cite{Adnktop} report ${\geqslant}\,10^{
12}$ equivalence classes of ${\geqslant}\,10^{47}$ topology types of off-shell 
supermultiplets of  $N\leqslant32$ worldline supersymmetry.  The point of  investigating 
these topics is to gain a deeper understanding---and computationally faster criteria---of 
when two adinkras are equivalent representations and thus describe the same 
physics representations.

In this paper, we focus on understanding such equivalences from a different direction.
This is a case study, but has a direct generalization to the formalism wherein
supermultiplets are represented by $\rd\times\rd$ matrices, which encode the orbits
of the supersymmetry charges amongst the component fields.  In particular, we start 
with the simple case of the ${\cal GR}(2,2)$ ``Garden Algebra'' generated by two 
$2\times2$ matrices, and then move to ${\cal GR}(4,4)$, generated by four $4\times4$ 
matrices.  In both cases, we utilize a factorization of each Garden Algebra matrix 
into an element of the permutation group multiplied by a Boolean (sign) factor. 
This decomposition affords both an extremely concise notation for the Garden 
Algebra matrices, and so for the supersymmetry orbits within the corresponding 
supermultiplets, and  a new and computationally efficient classification of equivalences. 
The decomposition is also straightforwardly generalized to all ${\cal GR}(\rd,N)$ 
matrices and therefore to all ${\geqslant}\,10^{47}$ topology types of $N\leqslant32$ 
worldline supersymmetry.

We use a deterministic calculation method enabled by a \textsl{Mathematica$^{\sss\text{TM}}$}
code to generate all possible matrices in the solution space and find all possible sets of 
$4\times4$ matrix solutions that satisfy the conditions of a Garden Algebra.  In doing so, we 
analyze the solution space and the possible transformations that can act on the sets of the 
solution matrices and relate them to previously described operations that define 
equivalences of adinkras. We then can organize the solution space and analyze the 
equivalence classes on the matrix representations with respect to the physical equivalence 
of the so-represented supermultiplets.

The uncovered class structure also indicates a natural operation in the space of the
permutation group class structure imposed on adinkras that is remarkably similar to
the well-known Hodge star operation in cohomology theory.

We will give this operation the name of the `$*$-map' acting on the space 
of matrix solutions.  The presence of this operation is used to organize the 
solution space and define new ways to introduce equivalence classes on 
the matrix representations of adinkras.  

Some striking new features become apparent as a result of this current 
analysis.  

In a previous work \cite{Genomics}, the representations of off-shell 4D, $\cal 
N$ = 1 supersymmetry with the least numbers of fields, i.\ e.\ the chiral scalar 
supermultiplet, the vector supermultiplet, and the tensor supermultiplet were 
used to create their corresponding adinkras.  As adinkra graphs possess 
adjacency matrices, these can be used to define character-like quantities 
called `chromocharacters.'  The details on the specific reduction route used 
was presented in the discussion.  The chromocharacters for the chiral scalar 
adinkra obtained were found to be different for those of the vector adinkra.  
Similarly, the chromocharacters from the chiral scalar adinkra were found to 
be different for those of the tensor supermultiplet.  In turn, the chromocharacters 
for the vector adinkra were found to be the same as those of the tensor 
adinkra, so that the vector adinkra and tensor adinkra are not distinguished 
by chromocharacters.
 
Herein, we demonstrate that the $*$-map lifts this degeneracy!  In fact, under 
the action of the $*$-map, the vector multiplet adinkra is a singlet, while the 
chiral scalar and tensor multiplet adinkras are exchanged for each other.   This 
means that the fundamental representation space structure for off-shell 4D, 
$\cal N$ = 1 supersymmetry is uncannily similar to that of the familiar 
representations $\bf3$, or $\bf3^*$, and $\bf8$ of the $\textit{su}(3)$ algebra.
 However, there are also 
significant differences between the structure of the familiar $\textit{su}(3)$ algebra
and the organization and structure of ${\cal GR}(4,4)$, as we discuss in closing.

\section{Description of Adinkras}
\label{s3}
The formal definition of an adinkra can be found in previous works
 \cite{Auto,Adnktop,Genomics,Adinkras}. Here, we will present but a brief introduction. 

An adinkra is a graphical representation of a supersymmetric multiplet (supermultiplet)
where component fields depend solely on a temporal coordinate.  The graph satisfies certain
relations  among its elements. An adinkra has nodes, colored black for fermions and white 
for bosons. Each node is drawn at a height proportional to the engineering dimension of
the component field it depicts, so nodes of distinct color never appear at the same height, 
and the integrally spaced height levels are populated with black and white nodes, alternating.  
The links in an adinkra represent the orbits of the supersymmetry charges acting on the 
nodes, and so connect {\em {only}} nodes of opposite color. The links are colored distinctly, 
in correspondence with the $N$ distinct generators of the $N$-extended supersymmetry they
represent, \ie, the links in the adinkra form $N$ equivalence classes by color.  The links in an 
adinkra may also be solid (depicting a factor of $+1$) and dashed (depicting a factor of $-1$) in 
the supersymmetry action acting on fields; Refs.\cite{Adnktop} prove that this suffices. The adinkras also satisfy a closed path (cycle) rule such that any cycle of 4 links must have an odd 
number of negative signs (dashed links).

 So, each adinkra with $\rd$ white and $\rd$ black nodes connected by links of $N$ colors
depicts an off-shell supermultiplet with $\rd$ bosonic and $\rd$ fermionic component fields
connected by the acton of $N$ supersymmetries. The (sign-modified and color-filtered)
adjacency matrices of this graph satisfy a Clifford algebra-like condition, and this condition 
defines the corresponding ${\cal GR}(\rd,N)$ algebra \cite{rGR2} which later acquired 
the name ``Garden Algebra.''

In Ref.\cite{Genomics}, a review of six supersymmetric multiplets (off-shell versus on-shell 
were counted as inequivalent for the purposes of the study) was given in terms 
of the fields and the superspace covariant derivative. The resulting equations for the 
supersymmetric relations between fermionic and bosonic fields were condensed into 
a set of matrix equations. 

For example, the 1D, $N=4$ chiral multiplet\footnote{As its name connotes, this supermultiplet 
is related to the usual chiral scalar supermultiplet in four dimensions.} consists of the 
bosonic fields $\F_i$ (for $i = 1\dots4$ respectively these correspond to functions
$A,B,F,G$) and the fermionic fields $\J_\hk$ ($\hk=1\dots4$) with the superspace 
covariant derivatives $\rD_{\sss\rI}$ and time derivative $\vd_0=\frac{\rd}{\rd t}$.  A 
supersymmetric system of equations can be written as
\begin{align}
   \rD_{\sss\rI}\, \F_i ~&=~ i \,(\rL_{\sss\rI})_i{}^\hk \, \J_\hk,
 \label{chiD0E}\\
   \rD_{\sss\rI}\, \J_\hk ~&=~(\rR_{\sss\rI})_\hk{}^i\,\frac{\rd}{\rd t}\,\F_i.
\label{chiD0J}
\end{align}
The 1D (worldline) dimensional reduction of (all off-shell, it is believed, ) familiar supermultiplets 
can be cast in this format, and in each case, the matrices $(\rL_{\sss\rI})$ and $(\rR_{\sss\rI})$ 
encode the action of the  supercharges among the component fields. 
For the transformations~(\ref{chiD0J}) to close the $N$-extended worldline supersymmetry 
algebra without central charges,
\begin{equation}
  \big\{\, \rD_{\sss\rI} \,,\, \rD_{\sss\rJ} \,\big\} ~=~ 2i  \, \d_{\sss\rI\,\rJ} \, \vd_0,
 \label{e:SuSyD}
\end{equation}
the L- and R-matrices must satisfy the algebraic equations
\begin{subequations}
 \label{GR(d,N)}
\begin{align}
  (\rL_{\sss\rI})_i{}^\hj\,(\rR_{\sss\rJ})_\hj{}^k
  +(\rL_{\sss\rJ})_i{}^\hj\,(\rR_{\sss\rI})_\hj{}^k 
   &= 2\, \d_{\sss\rI\,\rJ}\, \d_i{}^k,
 \label{gt1}\\
  (\rR_{\sss\rI})_\hi{}^j\,(\rL_{\sss\rJ})_j{}^\hk
  +(\rR_{\sss\rJ})_\hi{}^j\,(\rL_{\sss\rI})_j{}^\hk 
   &= 2\, \d_{\sss\rI\,\rJ}\, \d_\hi{}^\hk,
 \label{gt2}
\end{align}
which define the Garden Algebra, ${\cal GR}(\rd,N)$. Note that the $\rI=\rJ$ case 
in these implies that $(\rR_{\sss\rI})=(\rL_{\sss\rI})^{-1}$ for each 
fixed value of the subscript I. Since these L- and R-matrices must map real 
bosonic component fields into real fermionic ones and back, we additionally 
require that
\begin{equation}
  (\rR_{\sss\rI})_\hj{}^k \d_{ik} = (\rL_{\sss\rI})_i{}^\hk \d_{\hj\hk},
  \quad\ie\quad
  (\rR_{\sss\rI}) = (\rL_{\sss\rI})^T,
 \label{gt3}
\end{equation}
so that each R-matrix is fully specified in terms of the corresponding L-matrix, 
which in turn satisfy
\begin{equation}
  (\rL_{\sss\rI})^T = (\rL_{\sss\rI})^{-1}
 \label{OrthoL}
\end{equation}
and so are orthogonal, real-valued matrices. In what follows, we use this to 
save space and specify the L-matrices, relying on Eq.~(\ref{gt3}) for the 
determination of the corresponding R-matrices.
\end{subequations}

As it turns out, for many of the best-known supermultiplets the L-matrices are 
{\em\/signed permutation\/} matrices\cite{EncyMath}: they have a single nonzero 
entry in every row and in every column, and the nonzero entries are $\pm1$; 
see Ref.\cite{Adnktop} for a proof of the last property. This implies that the given 
supermultiplet admits a basis of component fields (to which the matrices such 
as~(\ref{ChiL}) refer) such that every supercharge transforms every component 
field into precisely one other component field or its $\vd_0$-derivative.  While 
such supermultiplets are already surprisingly numerous (as cited in the introduction), 
they can be used as ``building blocks'' to construct considerably more complex 
supermultiplets\cite{NonAdinkras}.

Restricting our further considerations only to such signed permutation L-matrix 
solutions of the system~(\ref{GR(d,N)}), we note that all such matrices factorize
\begin{equation}
 (\rL_{\sss\rI})_i{}^\hk ~=~ 
     ({\cal S}^{\sss(\rI)})_i{}^\hl\, ({\cal P}_{\!\sss(\rI)})_\hl{}^\hk,
      \qquad \text{for each fixed }\rI=1,2,\dots,N.
\label{aas1}
\end{equation}
Here, the sign-matrix ${\cal S}^{\sss(\rI)}$ is a diagonal $\rd\times\rd$ matrix with 
only $\pm1$ entries on the diagonal, and  each ${\cal P}_{\!\sss(\rI)}$ is a matrix 
representation of a permutation of $\rd$ objects.

Writing the $i^\text{th}$ diagonal entry in the sign-matrix ${\cal S}^{\sss(\rI)}$ as 
$(-1)^{b_i}$ where $b_i=0,1$, we assemble the binary exponents $b_i$ into a 
$\rd$-bit binary ``word,'' which is the binary encoding of a natural number, 
$\opR_{\sss\rI}$, the ``sign-number.'' These completely and unequivocally 
encode the sign-matrix:
\begin{equation}
({\cal S}^{\sss(\rI)})_i{}^\hl
=\begin{bmatrix}
(-1)^{b_1} & 0 & 0 & \cdots\\
0 & (-1)^{b_2} & 0 & \cdots\\
0 & 0 & (-1)^{b_2} & \cdots\\
\vdots & \vdots & \vdots & \ddots\\
\end{bmatrix}
\quad\iff\quad
\Big(\opR_{\sss\rI} = \sum_{i=1}^\rd b_i\, 2^{i-1}\Big)_b
~~=~~
\underset{\SSS\text{(binary ``word'')}\atop
\SSS\text{(reversed)}}{[b_1b_2\cdots b_\rd]_2}
 \label{binSign}
\end{equation}
In turn, the $\rd\times\rd$ permutation matrices $({\cal P}_{\!\sss(\rI)})_\hl{}^\hk$ 
are precisely the standard (unsigned) adjacency matrices of the associated 
adinkra graph \cite{GrphThry}. Permutations of $\rd$ objects may be represented 
extremely compactly as a sequence of natural numbers that is reordered to 
indicate the permutation. For example, when d=4, we can write
\begin{equation}
  \begin{bmatrix}
   0 & 1 & 0 & 0\\
   0 & 0 & 1 & 0\\
   0 & 0 & 0 & 1\\
   1 & 0 & 0 & 0\\
  \end{bmatrix}~
  \iff\vev{2341},\qquad
  \begin{bmatrix}
   0 & 1 & 0 & 0\\
   1 & 0 & 0 & 0\\
   0 & 0 & 0 & 1\\
   0 & 0 & 1 & 0\\
  \end{bmatrix}~
  \iff\vev{2143},\quad
  \etc
 \label{e:P}
\end{equation}
The factorization~(\ref{binSign}) therefore separates the standard graph structure 
of the Adinkra encoded by the permutation matrix $({\cal P}_{\!\sss(\rI)})_\hl{}^\hk$, 
from the edge-dashing that encodes the negative signs in the 
system~(\ref{chiD0E})--(\ref{chiD0J}) encoded by the sign-matrix ${\cal S}^{\sss(\rI)}$.

Combining the notation\eq{binSign} with\eq{e:P} allows expressing the matrices 
of the ${\cal GR}(\rd,N)$ algebra in a compact form. For example, using d=4:
\begin{subequations}
 \label{e:SP}
\begin{alignat}9
  \hbox{\footnotesize$\begin{bmatrix}
   ~~0 & 1 & 0 & ~~0\\[-1pt]
   ~~0 & 0 & 1 & ~~0\\[-1pt]
   ~~0 & 0 & 0 &  -1\\[-1pt]
    -1 & 0 & 0 & ~~0\\[-1pt]
  \end{bmatrix}$} =
  \hbox{\footnotesize$\begin{bmatrix}
   1 & 0 & ~~0 & ~~0\\[-1pt]
   0 & 1 & ~~0 & ~~0\\[-1pt]
   0 & 0 &  -1 & ~~0\\[-1pt]
   0 & 0 & ~~0 &  -1\\[-1pt]
  \end{bmatrix}$}
  \hbox{\footnotesize$\begin{bmatrix}
   0 & 1 & 0 & 0\\[-1pt]
   0 & 0 & 1 & 0\\[-1pt]
   0 & 0 & 0 & 1\\[-1pt]
   1 & 0 & 0 & 0\\[-1pt]
  \end{bmatrix}$}
   &=(12)_b\vev{2341}
  &&=[0011]_2\vev{2341}
  &&= \vev{23\bar4\bar1},\\
  \hbox{\footnotesize$\begin{bmatrix}
   0 & 1 & ~~0 & ~~0\\[-1pt]
   1 & 0 & ~~0 & ~~0\\[-1pt]
   0 & 0 & ~~0 &  -1\\[-1pt]
   0 & 0 &  -1 & ~~0\\[-1pt]
  \end{bmatrix}$} =
  \hbox{\footnotesize$\begin{bmatrix}
   1 & 0 & ~~0 & ~~0\\[-1pt]
   0 & 1 & ~~0 & ~~0\\[-1pt]
   0 & 0 &  -1 & ~~0\\[-1pt]
   0 & 0 & ~~0 &  -1\\[-1pt]
  \end{bmatrix}$}
  \hbox{\footnotesize$\begin{bmatrix}
   0 & 1 & 0 & 0\\[-1pt]
   1 & 0 & 0 & 0\\[-1pt]
   0 & 0 & 0 & 1\\[-1pt]
   0 & 0 & 1 & 0\\[-1pt]
  \end{bmatrix}$}
   &=(12)_b\vev{2143}
  &&=[0011]_2\vev{2143}
  &&= \vev{21\bar4\bar3},
\end{alignat}
\end{subequations}
and so on; the direct correspondence between the bars in the (right-most) notation 
for elements of the signed permutation group, the sign-factor $({\cal S}^{\sss(\rI)})$ 
in the factorization\eq{aas1} and the (reversed) binary ``word'' $[b_1b_2b_3b_4]_2
$ should be evident. 

These observations about the factorization~(\ref{aas1}) allow us to easily count the 
number of possible signed permutation matrices from which certain $N$-tuples 
satisfy the Garden Algebra conditions~(\ref{GR(d,N)}), with the result 
\bea
  \# (\rL_{\sss\rI}) ~=~ 2^d \, d!~.
\label{aas3}
\eea
Indeed, this is the dimension of the Coxeter group $BC_4=S_2\wr S_4$, the finite 
multiplicative group formed by all $4\times4$ signed permutation matrices\cite{EncyMath}. 
We are thus exploring possible embeddings ${\cal GR}(4,4)\subset BC_4$, and the 
corresponding $4\times4$ matrix realizations of ${\cal GR}(4,4)$ within those of 
$BC_4$.

With such a large (combinatorially growing with $\rd$) number of matrices to choose 
from, it is imperative to determine effective criteria for partitioning these signed permutation 
matrices into equivalence classes, and counting those equivalence classes effectively. 
Furthermore, we must determine the equivalence relations to faithfully correspond to 
the physical equivalence of the supermultiplets encoded by these L-matrices and the 
system of superdifferential equations~(\ref{chiD0E})--(\ref{chiD0J}).

This ``physical'' notion of equivalence enters in determining whether two adinkras and 
the corresponding worldline (1D) supermultiplets are dimensional reductions of the 
same or of different supermultiplets in higher-dimensional spacetimes.  Owing to this, 
the definition of `equivalence' itself is a subtle one.   More explicitly, the problem is 
how to determine if two adinkras are describing the same supersymmetric multiplet 
from the point of view of the physics---both on the 1D worldline, but more importantly, 
in the higher-dimensional spacetime.

We now turn to study several cases, using a \textsl{Mathematica$^{\sss\text{TM}}$} 
code to investigate the space of all sign permutation L-matrices as solutions of the 
appropriate system~(\ref{GR(d,N)}). In the next section, we thus analyze the solution 
space of ${\cal GR}(2,2)$, and then turn to the more interesting case of ${\cal GR}(4,4)$. 

\section{Sample Case of the  $\cg\car(2,2)$ Garden Algebra }
\label{s:GR(2,2)}
One of the simplest cases is d = 2, $N=2$ with two supersymmetric pairs of partners. 
A couple of additional rules  are apparent in this case:

\begin{enumerate}
  \item Every node has exactly $N=2$ links corresponding to the number of 
     different supersymmetric  operators that can act on the field associated with that node.
  \item Every closed path (cycle) in the adinkra must have an odd number of minus
     links in its path. A link can be solid (positive) or dashed (negative).
  \item Every node can only have one unique link to a supersymmetric partner field, \ie,
     the supermultiplet has signed permutation L-matrices and  is depicted by an adinkra.
\end{enumerate}

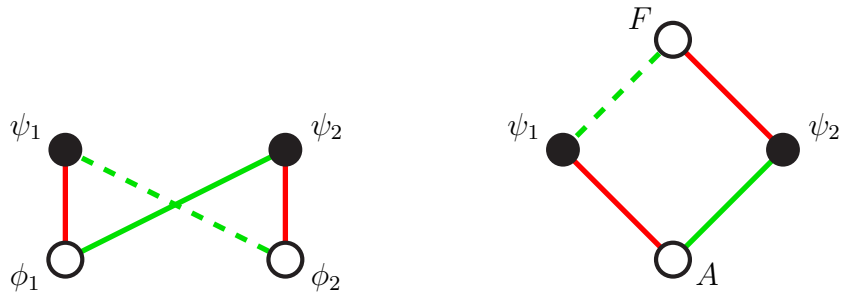
\begin{figure}[htbp]
$$
 \vCent{
  \begin{picture}(40,40)
   \put(2,2){\includegraphics[width=36mm]{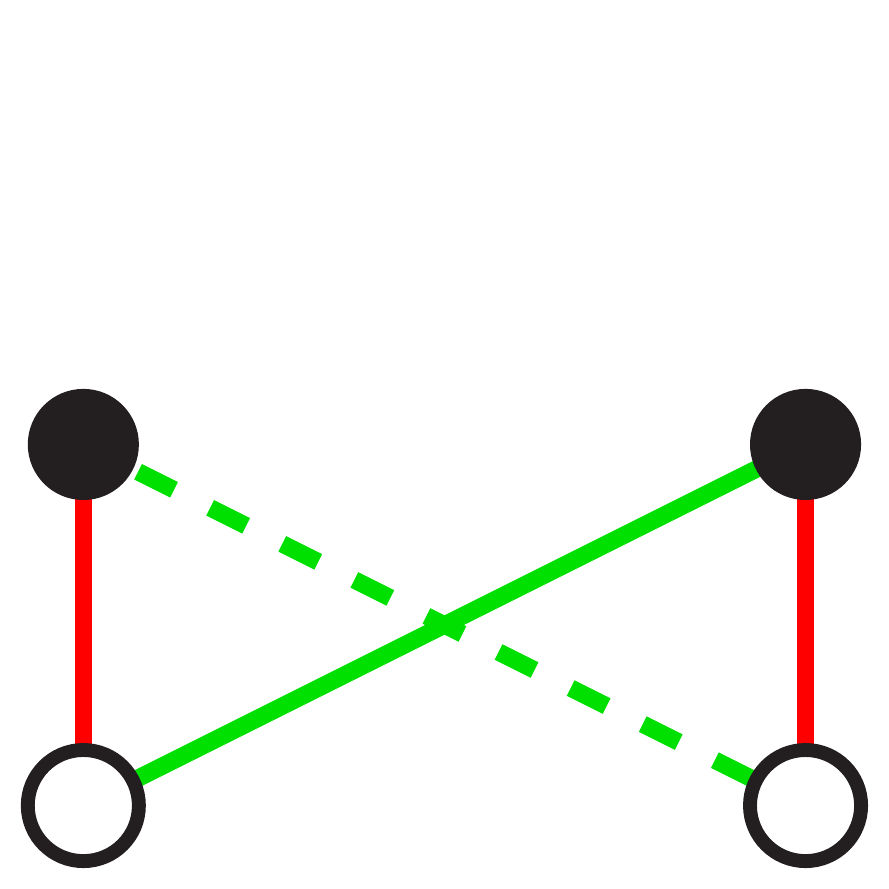}}
    \put(-2,2){$\f_1$}
    \put(-2,22){$\j_1$}
    \put(38,2){$\f_2$}
    \put(38,22){$\j_2$}
  \end{picture}}
 \qquad\qquad\qquad
 \vCent{
  \begin{picture}(40,40)
   \put(2,2){\includegraphics[width=36mm]{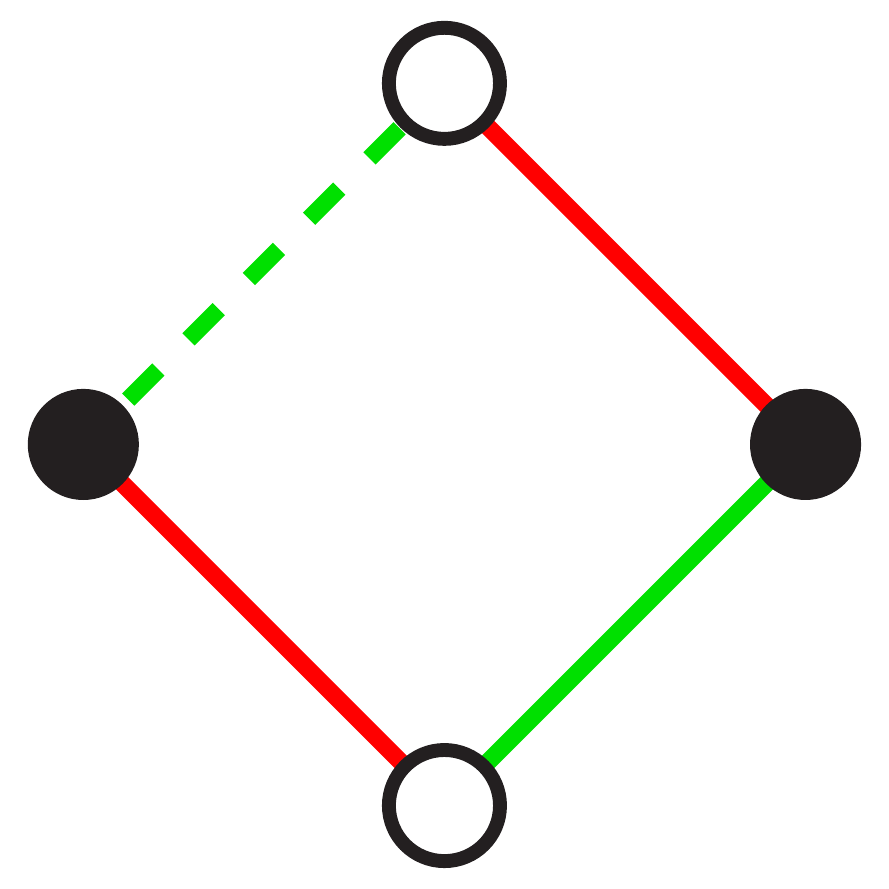}}
    \put(23,2){$A$}
    \put(-2,22){$\j_1$}
    \put(14,36){$F$}
    \put(38,22){$\j_2$}
  \end{picture}}
$$ \caption{Two $N=2$ adinkras, depicting two supermultiplets}
 \label{f:1}
\end{figure}

These rules only allow two basic types of adinkras at $\rd=2$, $N=2$, which may 
be called the bow-tie or the diamond, respectively.  The graphical representation 
of these are shown below with the bow-tie adinkra shown to the left and the diamond 
adinkra shown to the right in Figure~\ref{f:1}.

The first adinkra has two pairs of component fields: two bosonic fields $(\f_1,\f_2)$ 
with the same engineering dimension $[\f_1]=[\f_2]$, and two fermionic fields ($\j_1,
\j_2$) also of the same engineering dimension $[\j_1]=[\j_2]$. Also, $[\j_\hk]=[\f_i]{
+}\frac12$, for $i=1,2$ and $\hk=1,2$.  The second adinkra consists of a scalar field 
($A$), two fermions $(\j_1,\j_2)$ of the same engineering dimension, and a second 
scalar field $F$. The engineering dimensions now satisfy
$[F]{-}\frac12=[\j_1]=[\j_2]=[A]{+}\frac12$.

No other {\em {inequivalent}} adinkra can be constructed to satisfy the rules, except 
the fermion$\,{\iff}\,$boson flip of these two.  It may be better said that any other $N=
2$ adinkra is equivalent to one of these, up to certain equivalence operations. The 
known list of automorphisms acting on these graphs include:
\begin{description}
 \item[Edge-Color Swap:] Renaming the supercharges, \ie, swapping red $\leftrightarrow$ green
 \item[Dashing Flip:] `Flipping' solid links for dashed ones and vice-versa, while preserving
	 an odd number of dashed links
 \item[Node Swap:] Renaming the nodes variable at the same fixed height ($\phi_1 
	\leftrightarrow \phi_2$, $\psi_1 \leftrightarrow \psi_2$)
 \item[Node Sign Flip:] Changing the signs of some fields/nodes ($+\phi \leftrightarrow -\phi$) 
 \item[Klein Flip:] swapping the color of all nodes white $\iff$ black, \ie, swapping bosons $\iff$ fermions
\end{description}
The first two of these correspond to outer automorphisms acting on the supercharges.  
The next two correspond to inner automorphisms acting on the fields of the representation.  
The final one corresponds to a Klein transformation that exchanges bosons for fermions 
and {\em\/vice versa\/} throughout the supermultiplet.

We use the shorthand notation introduced by the factorization\eq{aas1} to describe
the 2 L-matrices for this case. From\eq{aas1}, the shorthand notation has a permutation 
factor ${\cal P}_{\!\sss(\rI)}$ and sign factor, ${\cal S}^{\sss(\rI)}$. We start we the 
permutation part: There exist only 2 permutations of two objects: $\vev{12}$ (the 
identity) and $\vev{21}$ (the swap).

In turn and up to the overall sign, there exist only two possible sign-matrices ${\cal 
S}^{\sss(\rI)}$:
\begin{alignat}9
 ({\cal P}_{\!\sss(1)})^{}_\hi{}^\hk
 &=\begin{bmatrix}
     ~1 & 0~ \\ ~0 & 1~
   \end{bmatrix}&
 &=\begin{bmatrix}
     ~(-1)^0 & 0~ \\ ~0 & (-1)^0~
   \end{bmatrix}&
  &~~\iff~~(0)_b&
  &=[00]_2~,\\
 ({\cal P}_{\!\sss(2)})^{}_\hi{}^\hk
 &=\begin{bmatrix}
     ~1 & 0~ \\ ~0 & -1~
   \end{bmatrix}&
 &=\begin{bmatrix}
     ~(-1)^0 & 0~ \\ ~0 & (-1)^1~
   \end{bmatrix}&
  &~~\iff~~(2)_b&
  &=[01]_2~,
\end{alignat}
the overall sign-flipped ones corresponding to $(1)_b=[10]_2$ and $(3)_b=[11]_2$.

We can quickly analyze the $N=2$ case by hand. Using the notation\eqs{chiD0E}{chiD0J} 
and restricting ourselves to the left-hand side adinkra in Figure~\ref{f:1}, we read off the 2 
$(=N)$ $\rL$-matrices:
\begin{alignat}9
 \C1{\rD_1}
    \begin{bmatrix}
      \f_1\\ \f_2
    \end{bmatrix}
 &=i\C1{\begin{bmatrix}
      1 & 0\\ 0 & 1
    \end{bmatrix}}
    \begin{bmatrix}
      \j_1\\ \j_2
    \end{bmatrix},&\qquad
   (\C1{\rL_1})^{}_i{}^\hk 
 &= \begin{bmatrix} 1&0\\ 0&1 \end{bmatrix}&
 &\overset{\text{(\ref{e:SP})}}{\longleftrightarrow}\vev{12}~;\\
 \C2{\rD_2}
    \begin{bmatrix}
      \f_1\\ \f_2
    \end{bmatrix}
 &=i\C2{\begin{bmatrix}
      ~~0 & 1\\ -1 & 0
    \end{bmatrix}}
    \begin{bmatrix}
      \j_1\\ \j_2
    \end{bmatrix},&\qquad
   (\C2{\rL_2})^{}_i{}^\hk 
 &= \begin{bmatrix} ~~0&1\\ -1&0 \end{bmatrix}&
 &\overset{\text{(\ref{e:SP})}}{\longleftrightarrow}\vev{2\bar1}~;
\end{alignat}

The corresponding words are $\rL_1 = (0)_{b}\vev{12}=\vev{12}$ and $\rL_2 
= (2)_{b}\vev{21}=\vev{2\bar1}$.  The total number of L-matrices is then $2^2\,2!=8$. 
Removing the overall minus sign redundancy, we have 4 matrices
\begin{equation}
  \vev{12}=\begin{bmatrix}1&0\\0&1\end{bmatrix},\quad
  \vev{1\bar2}=\begin{bmatrix}1&~~0\\0&-1\end{bmatrix},\quad
  \vev{21}=\begin{bmatrix}0&1\\1&0\end{bmatrix},\quad
  \vev{2\bar1}=\begin{bmatrix}~~0&1\\-1&0\end{bmatrix}.
\end{equation}
All other L-matrices can be generated by an overall minus sign.  It will prove 
useful to note that these matrices can be written in terms of Pauli matrices.  
Specifically, for the adinkras shown above we find
 ${\cal S}^{\sss(\rI)} = \{ \Ione_2, {\bm\s}^3 \}$ and
 ${\cal P}_{\!\sss(\rI)} = \{ \Ione_2,  {\bm \s}^1\}$, for $\rI=1,2$.
 
\section{Analysis of Adinkrizable Solutions in the Space of $\text{d} = 4$, $N=4$ Adinkras}
\label{AnalysisSoln}
We now turn the the case of the d = $4$, $N=4$ Garden Algebra adinkraic 
representations.

The matrices are generated algorithmically in two steps. The first step is to 
create the individual unsigned $4\times 4$ permutation matrices. Because 
of the assumptions made above, the matrices can be generated from a 
permutation of 4 objects giving $4!=24$ matrices that represent the elements 
of the permutation group $S_4$. The next step is to introduce all possible 
combinations of minus signs to generate all possible sets of Garden Algebra 
matrices.  This is done by creating a set of $4\times 4$ diagonal sign-matrices\eq{binSign}, 
of which there are $2^4=16$. Therefore, there exist a total of $4! \times 2^4=384$ 
product matrices\eq{aas1}.

We again use the shorthand notation developed in the last section to describe 
these matrices.  From\eq{aas1}, the shorthand notation has a permutation factor 
${\cal P}_{\!\sss(\rI)}$ and sign-factor, ${\cal S}^{\sss(\rI)}$. We denote these as 
before, in\eq{e:P} and\eq{binSign}, respectively, and then combine them as 
done in\eq{e:SP}.

For example, the L-matrices for the chiral multiplet as given in Ref.\cite{Genomics} 
may be decomposed as:
\begin{subequations}
 \label{ChiL}\vspace*{-5mm}
\begin{alignat}9
   &\qquad\qquad(\rL_{\sss\rI})^{}_i{}^{\hat k}&
   &=\quad\qquad({\cal C}^{\sss(\rI)})^{}_i{}^\hl
     \quad~~{\times}~~\quad({\cal P}_{\!\sss(\rI)})^{}_\hl{}^\hk&
   &=\makebox[0pt][l]{$(\bm{\cal R}_{\sss\rI})_b\vev{p_1p_2p_3p_4}$}
   \nn\\*[1mm]\hline\noalign{\nobreak\vglue1mm}
 (\rL_1)^{}_i{}^{\hat k} 
   &= \begin{bmatrix} 1&0&0&0\\ 0&0&0&-1\\ 0&1&0&0\\ 0&0&-1&0 \end{bmatrix}&
   &= \begin{bmatrix} 1&0&0&0\\ 0&-1&0&0\\ 0&0&1&0\\ 0&0&0&-1 \end{bmatrix}
       \begin{bmatrix} 1&0&0&0\\ 0&0&0&1\\ 0&1&0&0\\ 0&0&1&0 \end{bmatrix}&
   &=(10)_b\vev{1423}&
   &=\vev{1\bar42\bar3}~;
 \\
 (\rL_2)^{}_i{}^{\hat k}
   &= \begin{bmatrix} 0&1&0&0\\ 0&0&1&0\\ -1&0&0&0\\ 0&0&0&-1 \end{bmatrix}&
   &= \begin{bmatrix} 1&0&0&0\\ 0&1&0&0\\ 0&0&-1&0\\ 0&0&0&-1 \end{bmatrix}
       \begin{bmatrix} 0&1&0&0\\ 0&0&1&0\\ 1&0&0&0\\ 0&0&0&1 \end{bmatrix}&
   &=(12)_b\vev{2314}&
   &=\vev{23\bar1\bar4}~;
 \\
 (\rL_3)^{}_i{}^{\hat k}
   &= \begin{bmatrix} 0&0&1&0\\ 0&-1&0&0\\ 0&0&0&-1\\ 1&0&0&0 \end{bmatrix}&
   &= \begin{bmatrix} 1&0&0&0\\ 0&-1&0&0\\ 0&0&-1&0\\ 0&0&0&1 \end{bmatrix}
       \begin{bmatrix} 0&0&1&0\\ 0&1&0&0\\ 0&0&0&1\\ 1&0&0&0 \end{bmatrix}&
   &=(6)_b\vev{3241}&
   &=\vev{3\bar2\bar41}~;
 \\
 (\rL_4)^{}_i{}^{\hat k}
   &= \begin{bmatrix} ~0&~0&~0&~1\\ 1&0&0&0\\ 0&0&1&0\\ 0&1&0&0 \end{bmatrix}&
   &= \begin{bmatrix} 1&0&0&0\\ 0&1&0&0\\ 0&0&1&0\\ 0&0&0&1 \end{bmatrix}
       \begin{bmatrix} 1&0&0&0\\ 0&0&0&1\\ 0&1&0&0\\ 0&0&1&0 \end{bmatrix}&
   &=(0)_b\vev{4132}&
   &=\vev{4132}~.
\end{alignat}
\end{subequations}
So the sign-numbers of the L-matrices shown here are $(10)_b$, $(12)_b$, 
$(6)_b$, and $(0)_b$; they pertain to the 1D dimensional reduction of the 
chiral supermultiplet.

By representing the nodes by letters (greek for fermions and latin for bosons), we can see the effect
of the $\rL{}_{\sss\rI}$ and ${\rm R}{}_{\sss\rI}$ matrices directly. We let $(abcd)$ represent a vector
of the 4 bosons in a theory and $(\k \l \m \n)$ represent a vector of the 4 superpartner fermions and
ask how does the supersymmetric variation map the bosons into the fermions. For a theory with a L-matrix
 of $L_2 = \vev{23\bar1\bar4}$, we can apply \eq{chiD0E} to see the following:
\begin{alignat}4
 \rD_{\sss2}(a,b,c,d)^t
 &=i(\rL_2)^{}_i{}^{\hat k}\,(\k,\l,\m,\n)^t &\\
 &=i\,\vev{23\bar1\bar4}\,(\k,\l,\m,\n)^t
  &=i\,(\l,\m, -\k, -\n)^t, & \quad\ie,\\
  \rD_{\sss2} \begin{bmatrix}a\\ b\\ c\\ d\end{bmatrix}
  &=i\begin{bmatrix} 0&1&0&0\\ 0&0&1&0\\ -1&0&0&0\\ 0&0&0&-1 \end{bmatrix}
     \begin{bmatrix}\k\\ \l\\ \m\\ \n\end{bmatrix}
  &=i\begin{bmatrix}~~\l\\ ~~\m\\ -\k\\ -\n\end{bmatrix}~.&
\end{alignat}
What this means is for drawing the adinkra for this case, the boson $a$ is linked to fermion
$\l$, $b$ to $\m$, $c$ to $-\k$, and $d$ to $-\n$. \eq{chiD0J} calls for the use of the corresponding $(\rR_2)=(\rL_2)^t$ 
matrix, acting on the vector of bosons and giving the supersymmetry 
transformation of the fermion vector. This will be important when we 
discuss the equivalence of adinkras.

In terms of the decomposition\eq{aas1}, the three off-shell supermultiplets 
of Ref.\cite{Genomics} can be seen in table~\ref{t:0brane1} 
(see also appendix~\ref{a:123}), where we include the enantiomer numbers.
\begin{table}[!h]
$$
    \begin{array}{|c|c|c|c|c|c|c|} 
   \hline
 & \rL_1 & \rL_2 & \rL_3 & \rL_4 & n_c & n_t \\  \hline
CM & (10)_b \vev{1423} & (12)_b \vev{2314} &  (6)_b \vev{3241} & (0)_b \vev{4132} & 1 & 0
\\ \hline
VM & (10)_b \vev{2413} & (12)_b \vev{1324} &  (0)_b \vev{4231} & (6)_b \vev{3142} & 0 & 1
\\ \hline
TM & (14)_b \vev{1342} &  (4)_b \vev{2431} &  (8)_b \vev{3124} & (2)_b \vev{4213} & 0 & 1
\\ \hline
    \end{array}
$$
    \caption{The signed permutation element decomposition of L-matrices}
    \label{t:0brane1}
\end{table}

Ref.\cite{Genomics} also introduced 4 $\times$ 4 matrices denoted by ${\bm\a}_\rI$ and ${\bm\b}_\rI$. These may also factorized\eq{aas1} into a product of a sign- and a permutation factor:
\begin{equation}
\begin{array}{@{}c@{\,=\,}c@{\,=\,}c@{\,=\,}ccc@{\,=\,}c@{\,=\,}c@{\,=\,}c@{}}
 {\bm\a}^1 &{\bm\s}^2 \otimes {\bm\s}^1 &-i\,(12)_b\vev{4321}&-i\,\vev{43\bar2\bar1}, &&
 {\bm\b}^1 &{\bm\s}^1 \otimes {\bm\s}^2 &-i\,(10)_b\vev{4321}&-i\,\vev{4\bar32\bar1};
 \\*[2mm]
 {\bm\a}^2 &\Ione  \otimes {\bm \s}^2 &-i\,(10)_b\vev{2143}&-i\,\vev{2\bar14\bar3}, &&
 {\bm\b}^2 &{\bm\s}^2\otimes \Ione    &-i\,(12)_b\vev{3412}&-i\,\vev{34\bar1\bar2};
 \\*[2mm]
 {\bm\a}^3 &{\bm\s}^2 \otimes {\bm\s}^3 &-i\,(6)_b\vev{3412} &-i\, \vev{3\bar4\bar12}, &&
 {\bm\b}^3 &{\bm\s}^3 \otimes {\bm\s}^2 &-i\,(6)_b\vev{2143} &-i\, \vev{2\bar1\bar43}.
\end{array}
\end{equation}
and of course the 4 $\times$ 4 identity matrix correspond to
\be
   \Ione_4 ~=~ (0)_b\vev{1234}=\vev{1234}~.  
\ee
The significance of these observations is that both sets of four matrices
\be
  \{{\cal A}\} ~=~ \{\, \Ione_4\,,\, i\,{\bm\a}_\rI \,\}
   \quad\text{and}\quad
  \{{\cal B}\} ~=~ \{\, \Ione_4\,,\, i\,{\bm\b}_\rI \,\}
\label{albet}
\ee
also satisfy the conditions of \eq{GR(d,N)}.  

In turn, the set of matrices
\be
 \{{\cal M}\} ~=~ \{\, \Ione_4\,,\, i\,{\bm\a}_\rI\,,\, i\,{\bm\b}_\rI\,,\,
                        {\bm\a}_\rI{\bm\b}_\rJ \,\}
\label{M-set}
\ee
forms a complete basis for the expansion of all real $4\times4$ matrices.   It is also easy to establish that
under matrix transposition we find
\be
 \{{\cal M}\}^t ~=~ \{\, \Ione_4\,,\, -i\,{\bm\a}_\rI\,,\, -i\,{\bm\b}_\rI\,,\,
                          {\bm\a}_\rI{\bm\b}_\rJ \,\}.
\label{M-setT}
\ee
It is thus of interest to analyze these completely as representations of the signed permutation group. Our results are summarized in Table~\ref{t:0brane2}.
\begin{table}[!h]
$$
\begin{array}{|r@{\>=\>}r@{\>=\>}l|r@{\>=\>}r@{\>=\>}l|r@{\>=\>}r@{\>=\>}l|} 
         \hline
{\bm\a}_1 {\bm\b}_1 &  (6)_b \vev{1234} & \vev{1\bar2\bar34} & \rule{0pt}{2.5ex}
{\bm\a}_1 {\bm\b}_2 &  (0)_b \vev{2143} & \vev{2143} &
{\bm\a}_1 {\bm\b}_3 &  (5)_b \vev{3412} & \vev{\bar34\bar12}
 \\*[1mm]\hline
{\bm\a}_2 {\bm\b}_1 &  (0)_b \vev{3412} & \vev{3412}         & \rule{0pt}{2.5ex}
{\bm\a}_2 {\bm\b}_2 &  (9)_b \vev{4321} & \vev{\bar432\bar1} &
{\bm\a}_2 {\bm\b}_3 & (12)_b \vev{1234} & \vev{12\bar3\bar4}
 \\*[1mm]\hline
{\bm\a}_3 {\bm\b}_1 &  (3)_b \vev{2143} & \vev{\bar2\bar143} & \rule{0pt}{2.5ex}
{\bm\a}_3 {\bm\b}_2 & (10)_b \vev{1234} & \vev{1\bar23\bar4} &
{\bm\a}_3 {\bm\b}_3 &  (0)_b \vev{4321} & \vev{4321}
 \\*[1mm]\hline
\end{array}
$$
 \caption{The signed permutation element decomposition of the ${\bm\a}^{}_\rI{\bm\b}_\rJ$ matrices}
 \label{t:0brane2}
\end{table}

\section{Adinkras from Partitioning of the Permutation Group}
\label{Equiv}

In the last section, we described the process used to construct 384 matrices
to be taken as starting points for building all possible representations of the
$\cg\car(4,4)$ Garden Algebra.  Next these matrices were used to construct
all possible representations.

First a code was written to take all possible pairs of the 384 matrices and
identify the ones that satisfy the conditions in (\ref{gt2}) and (\ref{gt3}).
At this stage it was found that for any choice of the first member in the
pair there are sixteen other matrices that satisfy the required conditions.
The list of pairs was crossed-reference until finally, 1,536 sets of `tetrads' 
of four L-matrices each were identified.  

As the method used to generate the original 384 matrices involved the 
introduction of the boolean matrix factors, it was soon apparent that if the 
boolean factors were replaced by the identity matrix in all of the 1,536 
tetrad sets, they all could be identified with one of six partitions of the 
elements of the permutation group.  Each partition consists of a quartet
of permutation group elements.

Thus, it was discovered that the construction of all representations of the 
$\cg\car(4,4)$ Garden Algebra rests on a partitioning of the order four permutation 
group into six quartet sets.  These quartet sets will be given below.  This means 
for every permutation set, there were 1536/6 = 256 = $16^2$ boolean solutions. 
This is much smaller than the $16^4$ possible combinations of binary words that 
could have been solutions!  Looked at another way, we can summarize these 
results in one equation:
\begin{equation}
 \begin{aligned}
  16 \times (16\text{ binary words}) \times (6\text{ permutation quartets})
  &= 16 \times 384\text{ matrices}\\
  &= \,6,144\text{ matrices} ~=~ 1,536\text{ tetrads}~.
 \end{aligned}
\end{equation}
The 384 matrices correspond to a set of sixteen binary words assigned to
every element of the partitioned permutation group quartets.  The factor of 16 that 
multiplies the 384 represents the fact that given one solution consisting of four 
matrices, then any one of the four matrices may be replaced by its negative, 
to produce $2^4=16$ distinct solutions.  Stated differently, there are 
$\binom{384}{4}=891,881,376$ ways of selecting quartets of matrices from 
among the 384 elements of Coxeter's group $BC_4$; only 1,536 of these quartets
of signed permutation matrices furnish matrix representations of the
${\cal GR}(4,4)$ algebra, herein dubbed `tetrads.'

As we will make use of this partitioning later, it is useful to introduce some 
notation for the partitioned sets of quartets of elements of the permutation
group as
\begin{subequations}
 \label{six}
\begin{alignat}9
            &           & \rL_1\quad~ & \quad\rL_2  & \rL_3\quad~ & \quad\rL_4\nn\\
    \{CM \} &\equiv \{\,& \vev{1423},~& \vev{2314},~& \vev{3241},~& \vev{4132} \,\}~, 
    \label{six:CM} \\
    \{VM \} &\equiv \{\,& \vev{2413},~& \vev{1324},~& \vev{4231},~& \vev{3142} \,\}~, 
    \label{six:VM} \\
    \{TM \} &\equiv \{\,& \vev{1342},~& \vev{2431},~& \vev{3124},~& \vev{4213} \,\}~, 
    \label{six:TM} \\
\{{VM}_1 \} &\equiv \{\,& \vev{4123},~& \vev{1432},~& \vev{2341},~& \vev{3214} \,\}~, 
    \label{six:VM1} \\
\{{VM}_2 \} &\equiv \{\,& \vev{3421},~& \vev{4312},~& \vev{2134},~& \vev{1243} \,\}~, 
    \label{six:VM2} \\
\{{VM}_3 \} &\equiv \{\,& \vev{3412},~& \vev{4321},~& \vev{1234},~& \vev{2143} \,\}~, 
    \label{six:VM3}
\end{alignat}
\end{subequations}
and it is interesting to note that if we use a matrix representation for each of element
of the permutations indicated above, the following condition is satisfied:
\be
  \sum_{{\rI} = 1}^4 \, {\rm Tr} (\, {\cal P}_{\!\sss(\rI)} \,) ~=~ 4~,\quad
  \text{for all six sets\eq{six}.}
\ee
In a similar manner, for the sum the binary ``words''
 representing each set of sign-matrices with which the
 permutation matrices from Table~\ref{t:0brane1} close
 the ${\cal GR}(4,4)$ algebra, we find that
\be
   \sum_{{\rI} = 1}^4 \, ({{\cal R}_I})_b   ~=~ \textit{const.},
\ee
where the constant equals for all cases either 28 or 32, depending whether the L-matrices are replaced with their negatives.

 The L-matrices of the 1D (worldline) dimensional reduction of
 the familiar chiral multiplet belong to the $\{CM \}$ set, meaning
 that their permutation factors are listed in\eq{six:CM}. Similarly, the
 $\{VM \}$
set contains the vector multiplet solution, and the $\{TM \}$ set has the tensor multiplet 
solution.  (These are discussed in Appendix~\ref{a:123}.)
All the matrices in Table 2 occur in the sixth set.  We now have the 
possibility of a definition of equivalence class with respect to removing the signs 
from the L-matrices and considering only the (unsigned) permutation factors.

We can now change the question and ask what are the equivalence classes with 
respect to these permutation elements. We start with the first set which 
corresponds to the chiral multiplet. Because the elements are fixed inside this 
set, we can just focus on a single element in this set, $\vev{2314}$. For this 
element, there are 256 unique solution sets that solve \eq{GR(d,N)}. We can factor out 
16 sets of sets as being the same initial matrix $\vev{2314}$ multiplied by all 
possible $\pm1$ matrices $(\bm{\cal R}_n)_b$, for $n = 0\ldots15$. Keeping with the 
solution from the chiral multiplet, we are left with 16 sets of 4 matrices that all 
contain $(12)_b\vev{2314}$.

Looking at the sign codes of the other matrices in the solution sets, we find 
that there are only 6 sign codes. For $(12)_b\vev{2314}$, they are $(0)_b$,
$(5)_b$, $(6)_b$, $(9)_b$, $(10)_b$, and $(15)_b$. Upon closer inspection, we 
find that 3 are the exact opposite sign of the other three: $(0)_b = -(15)_b$,
$(5)_b = -(10)_b$, and $(6)_b = -(9)_b$. So finally, there are 3 sets of
sign-matrices and their negatives. If we look 
at the solution sets for the opposite sign of $(12)_b$, which is $(3)_b$, we 
find the exact same solution set. This accounts for all the possible differences 
between solution sets.

\subsection{Analysis of Transformations on Valise Adinkras
            \& The Permutation Basis  Elements}
A valise adinkra/supermultiplet\eqs{chiD0E}{chiD0J} is one that has all of the
bosons at one and the same level (have the same engineering dimension) and all
of the fermions at one and the same level, but of course
different from that of the bosons.  As noted in section~\ref{s3}, there are five 
types of transformations that can be done on one valise adinkra to obtain 
another valise adinkra.

Now that we have all the solutions (and a simple way to talk about them), we 
can clearly observe the effects that the adinkra transformations induce upon
the Garden Algebra matrices with regard to the relations to the elements
of the permutation group.

The benefit of the signed permutation representation is the simplicity of 
dealing with some of the combinatorics associated with adinkra transformations
described above.  For example, switching the labels of the 1st and 2nd nodes 
in an adinkra correspond to a transposition of the 1st and 2nd elements in the 
state, \ie, $(abcd) \rightarrow (bacd)$.

The first transformation, the edge-color swap, is simply a relabeling of the
adinkra. It is effectively 
relabeling the colors of the adinkra. In terms of the L-matrices, it is shifting 
the indices so $\rL_1 \to \rL_2$, \etc\ 
Thus, six distinct sets of the elements of the permutation group\eq{six} remain
distinct under this operation as sets of four $4\times4$ matrices.

The second transformation, the dashing flip, is equivalent to multiplying all
the $\rL_{\sss\rI}$ matrices by $-1$. Here again, the sign representations of
the $\rL_{\sss\rI}$ would change. However, because the original 
solution group contains both the original and $-1$ flipped versions of 
the sign representations, the solution set is effectively the same.
Thus the six distinct sets of the elements of the permutation set remain
distinct under this operation.

The third transformation, the node swap, is a relabeling of the fields at
a certain height. 
This corresponds to changing the order of the elements in one of the 
states, $(a_1a_2...a_i...a_j...) \rightarrow (a_1a_2...a_j...a_i...)$. The 
transformation is a permutation, $\cal P$, that can be applied to the
 $\rL_{\sss\rI}$ matrix or the other state vector, ($\m_1...)$. Applying 
it the $\rL{}_{\sss\rI}$ and more specifically to the permutation factor
of the representation, definitely changes the matrices and therefore the 
solution.  Thus the six distinct sets of the elements of the permutation 
group remain the same in number, but the solution sets are exchanged
under this operation.

The fourth transformation, the node sign flip, involves changing the sign
of one or more fields. This would involve a transformation of the sign
representation of the L-matrices.
This would not change the cycle part of the solution 
set but would change the sign part.  As shown above, all the possible 
sign combinations are already a part of the solution set.  Thus the six 
distinct sets of the elements of the permutation group remain
distinct under this operation.

The fifth listed transformation, the Klein flip, switches the bosons for
fermions and fermions for bosons.
Mathematically, this exchanges the vectors $\Phi_i$ 
and $\Psi_{\hat k}$ in equations \eqref{chiD0E} and \eqref{chiD0J}. To 
relate to the original formulation, we would have to switch the
 $\rL_{\sss\rI}$'s with the $\rR_{\sss\rI}$'s in the definitions. This 
is effectively mapping every matrix $\rL_{\sss\rI}$ to its transpose 
matrix $[(\rL_{\sss\rI})]^t$.  One might think that this 
does not change the solution sets. However upon inspection of all the 
permutation solution sets, we find something interesting. 

The Klein flip maps the first solution set (which contains $\vev{1432}$)
to the 5th solution set (which contains $\vev{1342}$, the transpose of
$\vev{1432}$ in $S_4$). This gives a relationship between the chiral
multiplet and the tensor multiplet.
All of the other solution sets, including the solution 
set for the vector multiplet, map back to themselves under the 
operation of taking the transpose of the L-matrices. However, it is only
the last set, $VM_3$, in which each of the four permutation factors in
the L-matrices is in fact symmetric. Thus, this is the only set which
maps to itself without requiring a compensating edge-color swap.

Of the five transformations in section~\ref{s3}, only the node swap
and the Klein flip may change the permutation factors of the solution set.
The Klein flip only changes two of the 
solution sets into each other. The node swap is the only 
one that changes the solution set completely. All the other 
transformations at most change the sign-factors inside a given solution set.

\subsection{A New Permutation Group Based Definition of Valise
            Adinkra Equivalence Classes and Implications}
We can take things a step further by analyzing {\em only} the node swap
and the Klein flip, and their effects in changing between permutation 
solution sets. The node swap can clearly change one of the 
6 solution sets into another depending on the reassignment of fields.  
We cannot define an equivalence class around this because the 
transformation makes no distinction between the solution sets: we 
can map any solution set into any other solution set with no loss of 
generality. We return to these transformations at the end of this section.

The Klein flip however breaks the solution sets into three definite classes:
\begin{enumerate}\itemsep=-3pt\vspace{-2mm}
 \item the two solution sets, $\{CM\}$ and $\{TM\}$, which are exchanged
       by the Klein flip;
 \item the three solution sets, $\{VM\}$, $\{VM_1\}$ and $\{VM_2\}$, which
       the Klein flip maps to \\  themselves, albeit up to some edge-color
       swapping;
 \item the one solution set, $\{VM_3\}$, which the Klein flip leaves fully
       unchanged.
\end{enumerate}
 Let us consider this situation further.
The action of transposition can also be considered directly on the
permutation factors, ${\cal P}_{\!\sss(\rI)}$.  If one begins with one element
of the permutation group $\cal A$, then the transposed element
${}^{\bm *}{\cal A}$ is simply the inverse, ${}^{\bm*}{\cal A}={\cal A}^{-1}$,
owing to Eq.\eq{OrthoL}.  Under the
action of this transposition operator, we find the sets satisfy
\begin{equation}
 \begin{array}{r@{\>=\>}l}
  {}^{\bm*}\{CM\}&\{TM^{\sss(c)}\}~,\\
  {}^{\bm*}\{TM\}&\{CM^{\sss(c)}\}~;\\
 \end{array}
 \qquad
 \begin{array}{r@{\>=\>}l}
  {}^{\bm*}\{VM\}  &\{VM^{\sss(c)}\}~,\\
  {}^{\bm*}\{VM_1\}&\{VM_1^{\sss(c)}\}~,\\
  {}^{\bm*}\{VM_2\}&\{VM_2^{\sss(c)}\}~,\\
 \end{array}
 \qquad
  {}^{\bm*}\{VM_3\}=\{VM_3\}~.
  \label{Hodge}
\end{equation}
The ``${}^{\sss(c)}$''superscript denoted that the L-matrices {\em\/within\/} the set
have been permuted.  

For the purposes of visualization, the space of 384 
matrices (representing the elements of the Coxeter group $BC_4$) can be illustrated
in terms of a pie chart where the sets $\{CM\}$, $\{TM\}$, $\{VM\}$, $\{VM_1\}$, $\{
VM_2\}$, and $\{VM_3\}$ each occupy one-sixth of the area.  
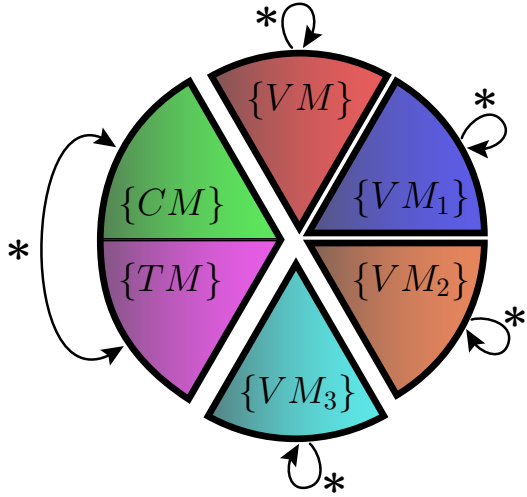
\begin{figure}[ht]
\begin{center}
\begin{picture}(70,60)
\put(0,-5){\includegraphics[height = 66\unitlength]{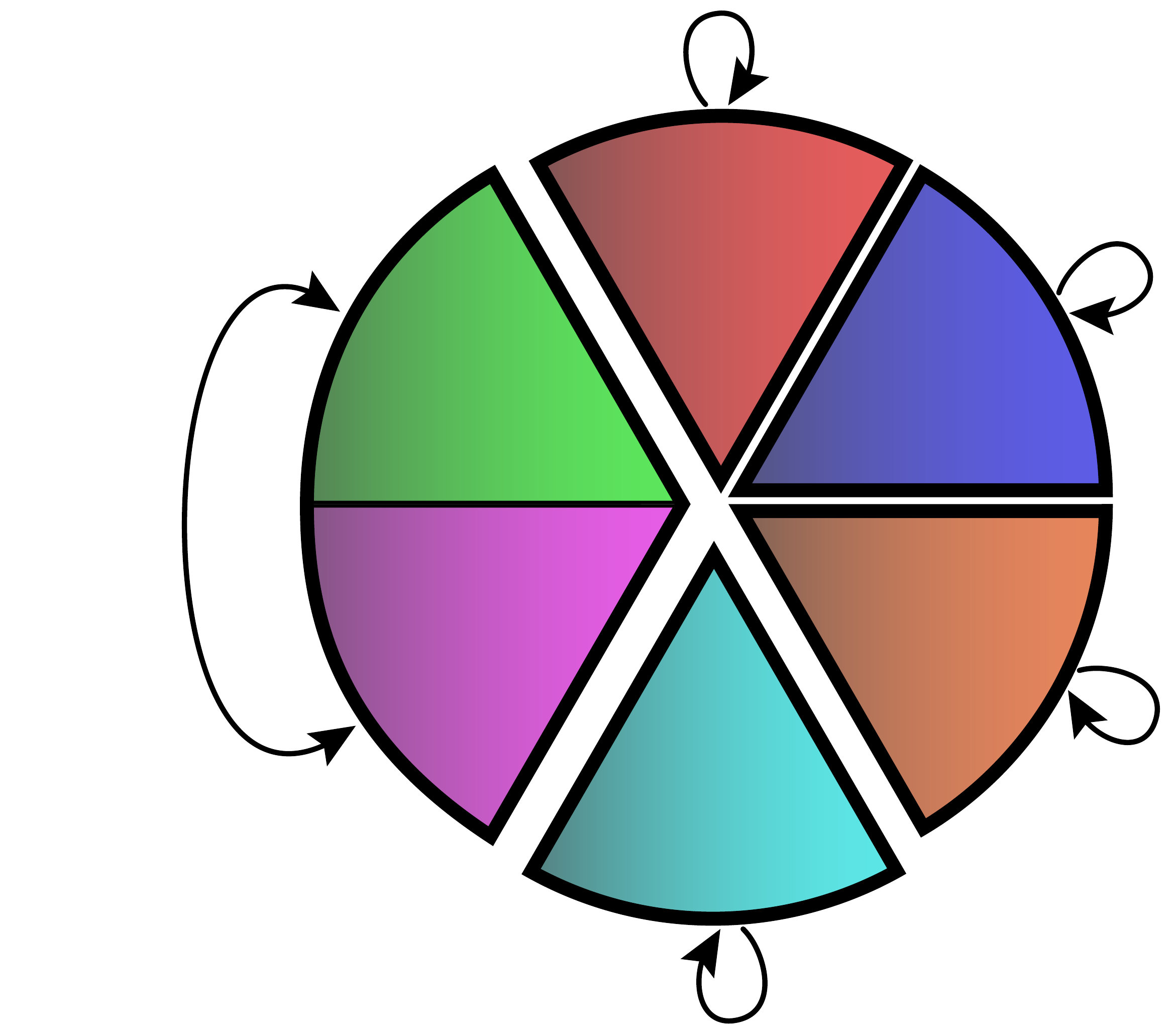}}
 \put(7,26){\Large$\bm*$}
 \put(40,57){\Large$\bm*$}
 \put(49,-5){\Large$\bm*$}
 \put(73,17){\Large$\bm*$}
 \put(69,46){\Large$\bm*$}
 \put(22,32){\large$\{CM\}$}
 \put(22,22){\large$\{TM\}$}
 \put(38,7){\large$\{VM_3\}$}
 \put(39,46){\large$\{VM\}$}
 \put(53,33){\large$\{VM_1\}$}
 \put(53,22){\large$\{VM_2\}$}
\end{picture}
\end{center}
\caption{Space of ${\cal GR}(4,4)$ matrices.}
\label{f:pie}
\end{figure}

The Klein flip operation acting on the adinkras is in 1--1 correspondence with
the $*$-operation acting on the elements of the both the signed and the unsigned
permutation groups, $BC_4$ and $S_4$. Therefore, the partitioning\eq{Hodge}
described also in the above enumeration as well as depicted in the pie-chart in
figure~\ref{f:pie} are all perfectly {\em\/intrinsic\/} to both $BC_4$ and $S_4$,
and so also to the complete solution set for the ${\cal GR}(4,4)$ matrix algebra.

In fact, this partitioning\eq{Hodge} induced by the action of the $*$-map also
follows from the elementary properties of the elements of the group of unsigned
permutations, $S_4$. Considering just the permutation factors of the $\{CM\}$,
$\{TM\}$ and $\{VM\}$ sets in table~\ref{t:0brane1} and the $\{VM_1\}$, $\{VM_2\}$,
and $\{VM_3\}$ sets in appendix~\ref{a:456}, we find:
\begin{enumerate}\itemsep=-3pt\vspace{-2mm}
 \item The $\{CM\}$ and $\{TM\}$ permutation factors are all order-3,
   \ie, their $3^\text{rd}$ power equals $\Ione_4$. Moreover, each $\{CM\}$
   permutation factor is the square of some $\{TM\}$ permutation factor, and
   also the other way around. This property pairs them, perfectly in line with
   the $*$-map pairing\eq{Hodge} also depicted in figure~\ref{f:pie}.
 \item The $\{VM\}$, $\{VM_1\}$ and $\{VM_2\}$ sets each have two
   permutation factors of order-2 and two of order-4, \ie, their $2^\text{nd}$
   and $4^\text{th}$ power equals $\Ione_4$, respectively.
 \item Only the $\{VM_3\}$ set has the identity $\Ione_4$ as one of the
   permutation factors, and the remaining three are of order-2, \ie, they
   square to $\Ione_4$.
\end{enumerate}

Considering next only the sign-matrices, represented by their sign-numbers, we find:
\begin{enumerate}\itemsep=-3pt\vspace{-2mm}
 \item The $\{CM\}$, $\{TM\}$ and $\{VM_3\}$ sets only use the odd permutations of
   the sign-number tetrads $\{(0)_b,(6)_b,(10)_b,(12)_b\}$ and
   $\{(2)_b,(4)_b,(8)_b,(14)_b\}$, a total of 24 sign-tetrads.
 \item Furthermore, each of these 24 sign-tetrads appears in two of the $\{CM\}$,
   $\{TM\}$ and $\{VM_3\}$ sets, none in all three. Stated differently, eight of
   the 24 sign-tetrads appear in $\{CM\}$ and $\{TM\}$, eight in $\{CM\}$ and
   $\{VM_3\}$, and the last eight in $\{TM\}$ and $\{VM_3\}$.\\[1mm]
\leavevmode\hglue0pt\kern-\leftmargin On the other hand,
 \item The $\{VM\}$, $\{VM_1\}$ and $\{VM_2\}$ sets only use the even permutations
   of the sign-number tetrads $\{(0)_b,(6)_b,(10)_b,(12)_b\}$ and
   $\{(2)_b,(4)_b,(8)_b,(14)_b\}$, a total of 24 sign-tetrads.
 \item Furthermore, each of these 24 sign-tetrads appears in two of the $\{VM\}$,
   $\{VM_1\}$ and $\{VM_2\}$ sets, none in all three. Stated differently, eight of
   the 24 sign-tetrads appear in $\{VM\}$ and $\{VM_1\}$, eight in $\{VM\}$ and
   $\{VM_2\}$, and the last eight in $\{VM_1\}$ and $\{VM_2\}$.
\end{enumerate}
This partitioning of the 48 sign-tetrads (all the permutations of
 $\{(0)_b,(6)_b,(10)_b,(12)_b\}$ and of $\{(2)_b,(4)_b,(8)_b,(14)_b\}$, taken up
to overall sign) is consistent with the partitioning\eq{Hodge} of the (unsigned)
permutations. Therefore, that the partitioning\eq{Hodge}, as depicted in
figure~\ref{f:pie} extends from the (unsigned) permutation group $S_4$ to the
full signed permutation group, $BC_4$, and thus also to the space of matrix
representations of ${\cal GR}(4,4)$ and the corresponding adinkras. Finally,
since adinkras faithfully depict 1D supermultiplets of $N$-extended supersymmetry
which admit a basis of component fields wherein each supercharge transforms each
component fields into another component field or its derivative, the same
partitioning also extends to these supermultiplets.

It is then highly suggestive to expect various different equivalence classes of
${\cal GR}(4,4)$ repre\-sen\-ta\-tions---such as those depicted in figure~\ref{f:pie}---to
in fact correspond to different supermultiplets.
 It has been shown in this paper that combinatorial factors are fixed with respect to the solutions of the 
Garden Algebra equations. There are 6 combinatorial sets of 4 matrices that 
form solution sets. There are fixed sets of sign factors that are related to those 
solutions. The underlying permutation representations are the basis of natural
equivalence  classes of the solutions under the $*$-map operation, of taking the transpose
matrix. 

Going back to \cite{Genomics}, we ask what are the implications of this definition 
of equivalence class based on the transpose matrix operation.  The vector 
multiplet as defined there turns up in the class\eq{six:VM} which is inert under the
action of matrix transposition. 
 Similarly, the chiral multiplet and tensor multiplet (as identified in Ref.\cite{Genomics}) turn up in the distinct pair of classes\eq{six:CM} and\eq{six:TM}, which are mapped into each other by the $*$-map, implemented as the matrix transposition operation on the L-matrices.
 Taking this as a hint, we may consider a mapping between the fields in the two multiplets (see appendix~\ref{a:123}) and we find that $A \iff \vf$ and $\j_a \iff \c_a$ by inspection. 
This would further imply 
that all the fields $B$, $F$, and $G$ of the chiral multiplet are mapped to the 
components\Ft{Recall that in the construction of any adinkra for a component 
gauge field, only the field components in the Coulomb gauge 
occur in an adinkra} $B_{i \, j}$ of the skew-symmetric tensor $B_{\mu \nu}$; compare figures~\ref{f:CM} and~\ref{f:TM}, and see table~\ref{t:Bench} in the appendix~\ref{a:123}.

This observation comes together beautifully with the structure seen in (\ref{Hodge})
if we identify the dual map defined on the elements of the permutation group with
a Hodge star-like map acting on the space of fields in the four dimensional field theory.
Under this duality, a chiral supermultiplet is replaced by a tensor supermultiplet
and vice-versa.  Furthermore under this duality, a vector supermultiplet maps 
into another vector supermultiplet.  All of these observations are consistent with
the equations seen in (\ref{Hodge}) and provides further support for the concept
 of ``SUSY holography''\cite[ called ``RADIO'' therein]{GR0}.

A final implication of using the diagram in figure~\ref{f:pie} in order to define ${\cal GR}(4,4)$ equivalence classes is that it implies restrictions on certain transformations identified in the work of Ref.\cite{Genomics}. There it was observed if one begins with a
set of matrices $\rL_{\sss\rI}$ that satisfy\eq{gt1}, (\ref{gt2}), and\eq{gt3}, then it is
possible to construct another such set ${\Hat {\rm L}}{}_{\bj I}$ that will also satisfy
these conditions where
\begin{align}
\Hat\rL_{\sss\rI} ~&=~ {\cal X}\, \rL_{\sss\rI} {\cal Y}~,\label{eq:E03}
\intertext{and where}
{\cal  X} \, ({\cal X})^t ~=~   ({\cal X})^t  \, {\cal  X} ~&=~
{\cal  Y} \, ({\cal Y})^t ~=~   ({\cal Y})^t  \, {\cal  Y} ~=~   \Ione~.
\label{eq:E04}
\end{align}
These last equations imply that ${\cal X}$ and ${\cal Y}$ are orthogonal $4\times4$ matrices, and for our present purposes may well be assumed to be (discrete) elements of the unsigned permutation subgroup of the (continuous) orthogonal group, $O(4)$. That is, these transformation matrices implement all possible node swaps within a supermultiplet, as defined in section~\ref{s:GR(2,2)}. This means that, through node-swaps, the L-matrices from every 4+4-component supermultiplet may be transformed so as to turn up in any one of the $*$-map equivalence classes\eq{Hodge} shown in figure~\ref{f:pie}. This then identifies the one remaining layer of relations between the structure uncovered by embedding ${\cal GR}(4,4)$ in Coxeter's signed permutation group $BC_4$, and the structure of the possible 4+4-component supermultiplets of $(N{=}4)$-extended supersymmetry on the worldline, and of ${\cal N}{=}1$ (simple) supersymmetry in 4D spacetime. We defer the study of these relations to a later effort.

\section{Conclusion}
\label{conclusions}
In this paper, we have established new results in defining a class structure
on Garden Algebras and their associated adinkras.  Though this work only concerned
the specific example of ${\cal GR}(4,4)$, it has wide implication far beyond
this example.  Any ${\cal GR}({\rm d},N)$ algebra in which ${\cal GR}(4,4)$ can
be embedded must inherit this equivalence class structure.  This definition
of equivalence classes is robust in the sense that it depends on intrinsic
properties of the Coxeter group $BC_4=S_2\wr S_4$ and therefore is independent
of the explicit representations chosen to write the L-matrices.  The definition
of adinkra equivalence described in this current work supersedes all previous
such assertions along these lines.

Our method of discovery was enabled by a \textsl{Mathematica$^{\sss\text{TM}}$} 
based search that allowed us to construct all 1,536 tetrads of L-matrices that
are monomial and satisfy the Garden Algebra conditions.  This permitted an
observation to be made that all such sets rely on a six-fold partitioning of the
permutation group $S_4$ into sets of four elements.
 These partitions were examined under the action of
a set of automorphism acting on the associated adinkras which where then
related to their action on elements of the permutation group.  One among the
these automorphisms, the $*$-map, was discovered to act within the Coxeter group
$BC_4$ akin to the well-known Hodge-star operator. Under the action of this
$*$-map operator, 256 of the tetrads (``$CM$'' in figure~\ref{f:pie}) are paired
with another 256 (``$TM$'' in figure~\ref{f:pie}), while the
remaining 1,024 of the tetrads are mapped to themselves (the various ``$VM$''
classes).

This partitions the 1,536 tetrads---the distinct matrix realizations of the
${\cal GR}(4,4)$ algebra---into the $*$-map pair of two 256-element equivalence
classes, and the $*$-map invariant equivalence class of 1,024 tetrads.

It is tempting to suggest that members in one half of this $*$-map pair of
equivalence classes provide an intrinsic definition of the chiral supermultiplet,
while their $*$-map images provide an intrinsic definition of the tensor
supermultiplet, and that the members of the remaining $*$-map invariant
equivalence class provide an intrinsic definition of the vector supermultiplet.

The fact that there may well exist an intrinsic definition of off-shell
supersymmetry representations based on the partitioning of the permutation
group under the action of a Hodge-star like operator, and that this seems to
dovetail precisely with the three known minimal representations (the chiral
multiplet, tensor multiplet, and vector multiplet) of four dimensional simple
supersymmetry, suggest the beginning of a theory of introducing
four-dimensional spin-bundles on adinkras.
 There appears promise in continuing this work and we look
forward to more enlightening results in the future.


\bigskip
\begin{flushright}
\parbox{90mm}{\raggedright\it\leavevmode{}\llap{``}%
 No human investigation can be called real science
 if it cannot be demonstrated mathematically.''\newline
 $~~~~$--- Leonardo da Vinci}
\end{flushright}

\bigskip
\paragraph{Acknowledgments}
This research has been supported in part by NSF Grant PHY-09-68854,
the J.~S. Toll Professorship endowment and the UMCP Center for
String \& Particle Theory.
 TH is grateful to the Physics Department of the Faculty of Natural Sciences of the University of Novi Sad, Serbia, for recurring hospitality and resources.

\clearpage
\vspace{5mm}
\appendix
\noindent{\large\sf\bfseries Appendices}
\section{Adinkras For The `Benchmark' CM, VM \& TM Sets}
\label{a:123}
The three sets of L-matrices identified in Table 1 are only significant in that they were identified
by a process of starting with the actual CM (chiral scalar super multiplet), VM (vector super multiplet),
and TM (tensor super multiplet) representations in four dimension and subjecting these 4D theories
to a reduction process \cite{Genomics}.  Of course, there is a large amount of arbitrariness in the 
choice of basis made for carrying out such calculations.  This implies that one could easily begin 
with the same starting point and end up with totally different L-matrices at the end.  So these are 
benchmarks in that they were the first explicitly derived set of L-matrices connected to known 4D 
supermultiplets.  The adinkras corresponding to these are given in the figures~\ref{f:CM}--\ref{f:TM} below.  
\begin{figure}[!ht]
 \begin{center}
  \begin{picture}(100,30)(2,0)
    \put(0,-10){\includegraphics[width = 100\unitlength]{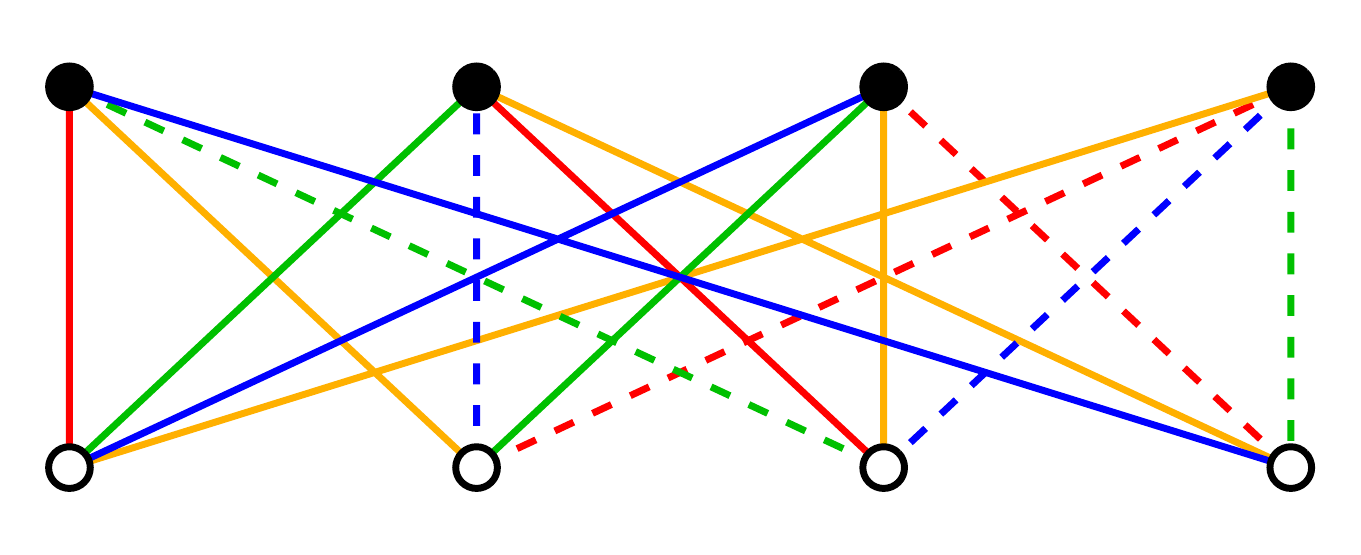}}
     \put(7,-7){$A$}
     \put(37,-7){$B$}
     \put(67,-7){$\int\!\rd\t\,F$}
     \put(97,-7){$\int\!\rd\t\,G$}
     \put(7,26){$-i\j_1$}
     \put(37,26){$-i\j_2$}
     \put(67,26){$-i\j_3$}
     \put(97,26){$-i\j_4$}
  \end{picture}
 \end{center}
\caption{CM adinkra}
\label{f:CM}
\end{figure}
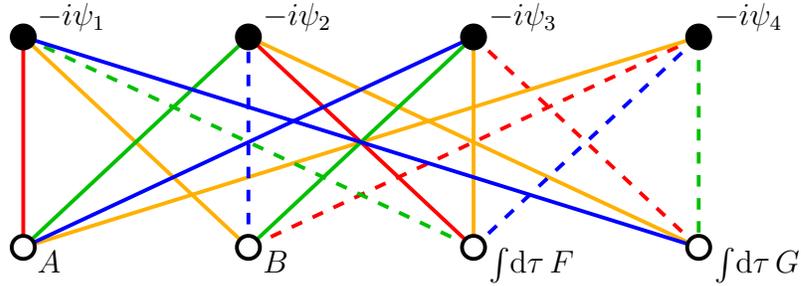
\begin{figure}[!ht]
 \begin{center}
  \begin{picture}(100,30)(2,0)
    \put(0,-10){\includegraphics[width = 100\unitlength]{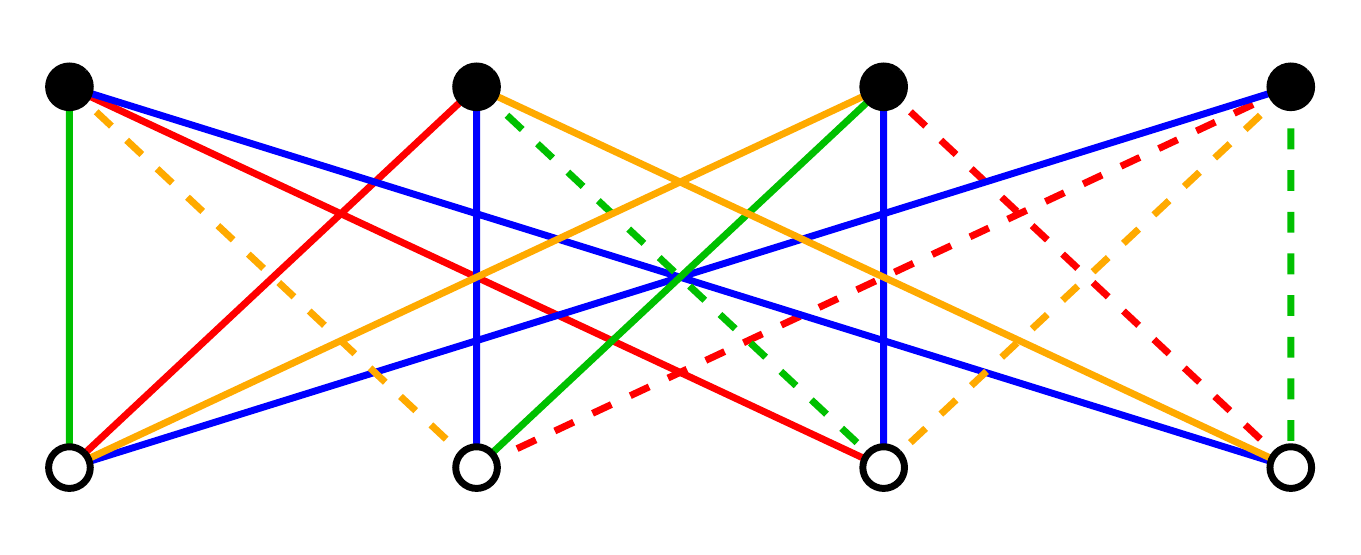}}
     \put(7,-7){$A_1$}
     \put(37,-7){$A_2$}
     \put(67,-7){$A_3$}
     \put(97,-7){$\int\!\rd\t\,d$}
     \put(7,26){$-i\l_1$}
     \put(37,26){$-i\l_2$}
     \put(67,26){$-i\l_3$}
     \put(97,26){$-i\l_4$}
  \end{picture}
 \end{center}
\caption{VM adinkra}
\label{f:VM}
\end{figure}
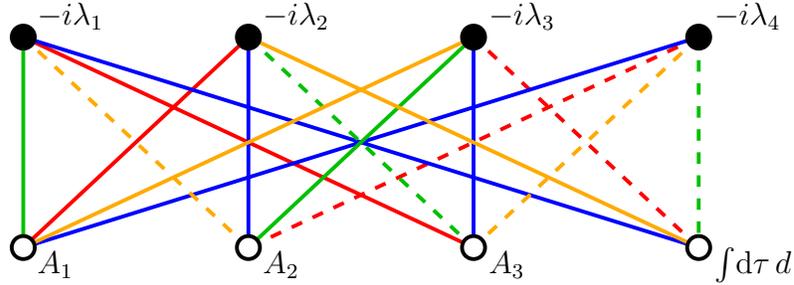
\begin{figure}[!ht]
 \begin{center}
  \begin{picture}(100,30)(2,0)
    \put(0,-10){\includegraphics[width = 100\unitlength]{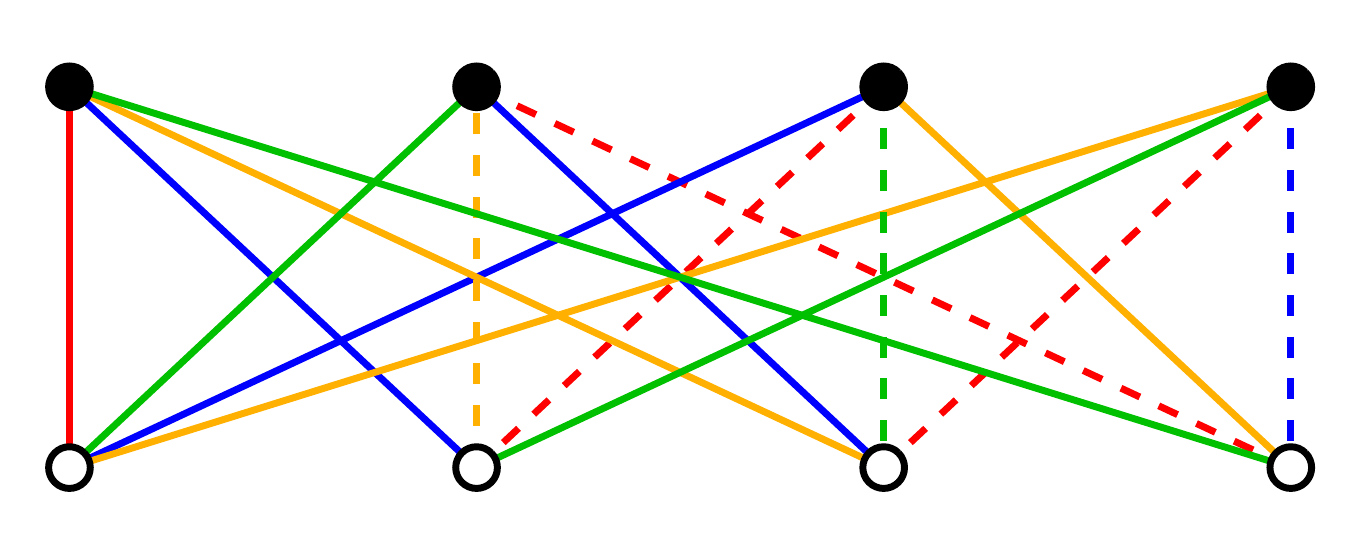}}
     \put(7,-7){$\vf$}
     \put(37,-7){$2B_{12}$}
     \put(67,-7){$2B_{23}$}
     \put(97,-7){$2B_{31}$}
     \put(7,26){$-i\c_1$}
     \put(37,26){$-i\c_2$}
     \put(67,26){$-i\c_3$}
     \put(97,26){$-i\c_4$}
  \end{picture}
 \end{center}
\caption{TM adinkra}
\label{f:TM}
\end{figure}
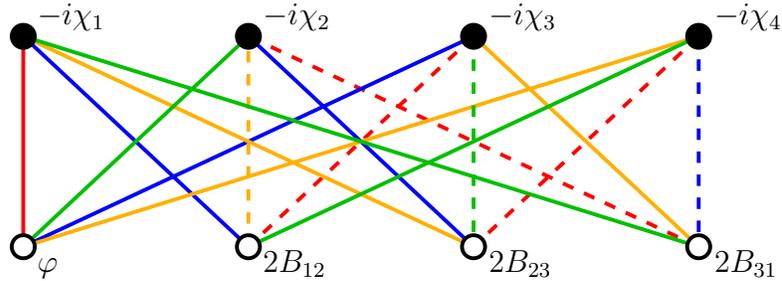

We note that the bosonic field variables and transformation laws for the CM set fields
have already been given in equations (\ref{e:SuSyD}).  As well, the L-matrices for
all three benchmark sets have been given in Table 1.  The only information remaining
in specifying the supersymmetric system equations is the relation of the bosonic
field variables $\F_i$ to the reduced field variables of the 4D systems.  This done in
table~\ref{t:Bench}; compare also with figures~\ref{f:CM},~\ref{f:VM} and~\ref{f:TM}.
\begin{table}[!h]
\small
$$
\begin{array}{|r|c|c|c|c||c|c|c|c|}
           \hline
 \text{Eqs.\eqs{chiD0E}{chiD0J}}
          & \F_1 & \F_2 & \F_3 & \F_4 \rule{0pt}{2.5ex}
          & \J_1 & \J_2 & \J_3 & \J_4
 \\*[1mm]  \hline
\{ CM  \} & A & B & \int\!\rd\t\,F & \int\!\rd\t\,G \rule{0pt}{2.5ex}
          &-i\j_1 &-i\j_2 &-i\j_3 &-i\j_4 
 \\*[1mm] \hline
\{ VM  \} & A_1 & A_2 &  A_1 & \int\!\rd\t\,d \rule{0pt}{2.5ex}
          &-i\l_1 &-i\l_2 &-i\l_3 &-i\l_4 
 \\*[1mm] \hline
\{TM \} & \varphi &  2\,B_{1\,2} &  2\,B_{2\,3} & 2\,B_{3\,1} \rule{0pt}{2.5ex}
          &-i\c_1 &-i\c_2 &-i\c_3 &-i\c_4 
 \\*[1mm] \hline
\end{array}
$$
\caption{Benchmark bosonic and fermionic fields}
\label{t:Bench}
\end{table}

\section{Garden Algebra Representatives From Three Permutation Sets}
\label{a:456}
A representative from the fourth, $\{VM_1\}$ set is given $(6)_b\vev{1432}$, $(3)_b\vev{2341}$, $(10)_b\vev{3214}$, 
and $(0)_b\vev{4123}$, which implies the following four matrices
\begin{equation}
 \begin{aligned}
  (\rL_1)^{}_i{}^\hk 
  &=\begin{bmatrix} 1&0&~~0&~~0\\ 0&0&~~0&-1\\0&0&-1&~~0\\ 0&1&~~0&~~0 \end{bmatrix}~,&\qquad
  (\rL_2)^{}_i{}^\hk
  &=\begin{bmatrix} 0&-1&~~0&0\\ 0&~~0&-1&0\\ 0&~~0&~~0&1\\ 1&~~0&~~0&0 \end{bmatrix}~, \\
  (\rL_3)^{}_i{}^\hk
  &=\begin{bmatrix} 0&~~0&1&~~0\\ 0&-1&0&~~0\\ 1&~~0&0&~~0\\ 0&~~0&0&-1 \end{bmatrix}~,&\qquad
  (\rL_4)^{}_i{}^\hk
  &=\begin{bmatrix} 0&0&0&1\\ 1&0&0&0\\ 0&1&0&0\\ 0&0&1&0\end{bmatrix}~.
\end{aligned}
 \label{e:VM1}
\end{equation}

A representative from the fifth, $\{VM_2\}$ set is given $(12)_b\vev{1243}$, $(9)_b\vev{2134}$, $(0)_b\vev{3421}$, 
and $(10)_b\vev{4312}$, which implies the following four matrices
\begin{equation}
 \begin{aligned}
  (\rL_1)^{}_i{}^\hk
  &= \begin{bmatrix} 1&0&~~0&~~0\\ 0&1&~~0&~~0\\ 0&0&~~0&-1\\ 0&0&-1&~~0 \end{bmatrix}~, &\qquad
  (\rL_2)^{}_i{}^\hk
  &=\begin{bmatrix} 0&-1&0&~~0\\ 1&~~0&0&~~0\\0&~~0&1&~~0\\ 0&~~0&0&-1 \end{bmatrix}~, \\
  (\rL_3)^{}_i{}^\hk
  &=\begin{bmatrix} 0&0&1&0\\ 0&0&0&1\\ 0&1&0&0\\ 1&0&0&0 \end{bmatrix}~, &\qquad
  (\rL_4)^{}_i{}^\hk
  &=\begin{bmatrix} 0&~~0&~~0&1\\ 0&~~0&-1&0\\ 1&~~0&~~0&0\\ 0&-1&~~0&0 \end{bmatrix}~.
\end{aligned}
 \label{e:VM2}
\end{equation}

A representative from the sixth, $\{VM_3\}$ set is given $(12)_b\vev{1234}$, $(5)_b\vev{2143}$, $(0)_b\vev{3412}$, 
and $(6)_b\vev{4321}$, which implies the following four matrices
\begin{equation}
 \begin{aligned}
  (\rL_1)^{}_i{}^\hk
  &=\begin{bmatrix} 1&0&~~0&~~0\\ 0&1&~~0&~~0\\0&0&-1&~~0\\ 0&0&~~0&-1 \end{bmatrix}~, &\qquad
  (\rL_2)^{}_i{}^\hk
  &= \begin{bmatrix} 0&-1&0&~~0\\ 1&~~0&0&~~0\\ 0&~~0&0&-1\\ 0&~~0&1&~~0 \end{bmatrix}~, \\
  (\rL_3)^{}_i{}^\hk
  &=\begin{bmatrix} 0&0&1&0\\ 0&0&0&1\\ 1&0&0&0\\ 0&1&0&0 \end{bmatrix}~, &\qquad
  (\rL_4)^{}_i{}^\hk
  &=\begin{bmatrix} 0&~~0&~~0&1\\ 0&~~0&-1&0\\ 0&-1&~~0&0\\ 1&~~0&~~0&0 \end{bmatrix}~.
\end{aligned}
 \label{e:VM3}
\end{equation}

\section{Enumeration of Signed Permutation Operator Decomposition For $\bm{\cal GR}(4,4)$ Sets}
\label{a:BC4}

The full list of 384 matrices that form the solution space of the ${\cal GR}(4,4)$ Algebra
is given here: each set of four matrices listed below next to each other in a row is one
such solution, called a ``tetrad'' in the text. However, the more concise ``bracket-overbar''
notation introduced in the text discussion of\eq{e:SP} is used. 

The complete list of L-matrices associated with the $\{CM\}$-set begins by giving the
listing
\begin{equation}
\begin{array}{cccc}
 \vev{1423} & \vev{23\bar1\bar4} & \vev{3\bar24\bar1} & \vev{4\bar1\bar32} \\
 \vev{1423} & \vev{2\bar3\bar14} & \vev{32\bar4\bar1} & \vev{4\bar13\bar2} \\
 \vev{1\bar423} & \vev{23\bar14} & \vev{3\bar2\bar4\bar1} & \vev{413\bar2} \\
 \vev{1\bar423} & \vev{2\bar3\bar1\bar4} & \vev{324\bar1} & \vev{41\bar32} \\
 \vev{142\bar3} & \vev{23\bar14} & \vev{3\bar241} & \vev{4\bar1\bar3\bar2} \\
 \vev{142\bar3} & \vev{2\bar3\bar1\bar4} & \vev{32\bar41} & \vev{4\bar132} \\
 \vev{1\bar42\bar3} & \vev{23\bar1\bar4} & \vev{3\bar2\bar41} & \vev{4132} \\
 \vev{1\bar42\bar3} & \vev{2\bar3\bar14} & \vev{3241} & \vev{41\bar3\bar2} \\
\end{array}
 \qquad\qquad
\begin{array}{cccc}
 \vev{14\bar23} & \vev{231\bar4} & \vev{3\bar2\bar4\bar1} & \vev{4\bar132} \\
 \vev{14\bar23} & \vev{2\bar314} & \vev{324\bar1} & \vev{4\bar1\bar3\bar2} \\
 \vev{1\bar4\bar23} & \vev{2314} & \vev{3\bar24\bar1} & \vev{41\bar3\bar2} \\
 \vev{1\bar4\bar23} & \vev{2\bar31\bar4} & \vev{32\bar4\bar1} & \vev{4132} \\
 \vev{14\bar2\bar3} & \vev{2314} & \vev{3\bar2\bar41} & \vev{4\bar13\bar2} \\
 \vev{14\bar2\bar3} & \vev{2\bar31\bar4} & \vev{3241} & \vev{4\bar1\bar32} \\
 \vev{1\bar4\bar2\bar3} & \vev{231\bar4} & \vev{3\bar241} & \vev{41\bar32} \\
 \vev{1\bar4\bar2\bar3} & \vev{2\bar314} & \vev{32\bar41} & \vev{413\bar2}
\end{array}
\end{equation}
and to obtain all the matrices associated with the $\{ CM \}$-set one simply
introduces a factor of $\pm$ in front of each matrix.

The complete list of L-matrices associated with the $\{VM\}$-set begins by giving the
listing
\begin{equation}
\begin{array}{cccc}
 \vev{1324} & \vev{2\bar4\bar13} & \vev{3\bar14\bar2} & \vev{42\bar3\bar1} \\
 \vev{1324} & \vev{24\bar1\bar3} & \vev{3\bar1\bar42} & \vev{4\bar23\bar1} \\
 \vev{132\bar4} & \vev{24\bar13} & \vev{3\bar1\bar4\bar2} & \vev{4\bar231} \\
 \vev{132\bar4} & \vev{2\bar4\bar1\bar3} & \vev{3\bar142} & \vev{42\bar31} \\
 \vev{1\bar324} & \vev{24\bar13} & \vev{314\bar2} & \vev{4\bar2\bar3\bar1} \\
 \vev{1\bar324} & \vev{2\bar4\bar1\bar3} & \vev{31\bar42} & \vev{423\bar1} \\
 \vev{1\bar32\bar4} & \vev{2\bar4\bar13} & \vev{31\bar4\bar2} & \vev{4231} \\
 \vev{1\bar32\bar4} & \vev{24\bar1\bar3} & \vev{3142} & \vev{4\bar2\bar31} \\
\end{array}
  \qquad\qquad
\begin{array}{cccc}
 \vev{13\bar24} & \vev{2\bar413} & \vev{3\bar1\bar4\bar2} & \vev{423\bar1} \\
 \vev{13\bar24} & \vev{241\bar3} & \vev{3\bar142} & \vev{4\bar2\bar3\bar1} \\
 \vev{13\bar2\bar4} & \vev{2413} & \vev{3\bar14\bar2} & \vev{4\bar2\bar31} \\
 \vev{13\bar2\bar4} & \vev{2\bar41\bar3} & \vev{3\bar1\bar42} & \vev{4231} \\
 \vev{1\bar3\bar24} & \vev{2413} & \vev{31\bar4\bar2} & \vev{4\bar23\bar1} \\
 \vev{1\bar3\bar24} & \vev{2\bar41\bar3} & \vev{3142} & \vev{42\bar3\bar1} \\
 \vev{1\bar3\bar2\bar4} & \vev{2\bar413} & \vev{314\bar2} & \vev{42\bar31} \\
 \vev{1\bar3\bar2\bar4} & \vev{241\bar3} & \vev{31\bar42} & \vev{4\bar231}
\end{array}
\end{equation}
and to obtain all the matrices associated with the $\{ VM \}$-set one simply
introduces a factor of $\pm$ in front of each matrix.

The complete list of L-matrices associated with the $\{TM\}$-set begins by giving the
listing
\begin{equation}
\begin{array}{cccc}
 \vev{1342} & \vev{2\bar43\bar1} & \vev{3\bar1\bar24} & \vev{42\bar1\bar3} \\
 \vev{1342} & \vev{24\bar3\bar1} & \vev{3\bar12\bar4} & \vev{4\bar2\bar13} \\
 \vev{13\bar42} & \vev{243\bar1} & \vev{3\bar1\bar2\bar4} & \vev{4\bar213} \\
 \vev{13\bar42} & \vev{2\bar4\bar3\bar1} & \vev{3\bar124} & \vev{421\bar3} \\
 \vev{1\bar342} & \vev{243\bar1} & \vev{31\bar24} & \vev{4\bar2\bar1\bar3} \\
 \vev{1\bar342} & \vev{2\bar4\bar3\bar1} & \vev{312\bar4} & \vev{42\bar13} \\
 \vev{1\bar3\bar42} & \vev{2\bar43\bar1} & \vev{31\bar2\bar4} & \vev{4213} \\
 \vev{1\bar3\bar42} & \vev{24\bar3\bar1} & \vev{3124} & \vev{4\bar21\bar3} \\
\end{array}
 \qquad\qquad
\begin{array}{cccc}
 \vev{134\bar2} & \vev{2\bar431} & \vev{3\bar1\bar2\bar4} & \vev{42\bar13} \\
 \vev{134\bar2} & \vev{24\bar31} & \vev{3\bar124} & \vev{4\bar2\bar1\bar3} \\
 \vev{13\bar4\bar2} & \vev{2431} & \vev{3\bar1\bar24} & \vev{4\bar21\bar3} \\
 \vev{13\bar4\bar2} & \vev{2\bar4\bar31} & \vev{3\bar12\bar4} & \vev{4213} \\
 \vev{1\bar34\bar2} & \vev{2431} & \vev{31\bar2\bar4} & \vev{4\bar2\bar13} \\
 \vev{1\bar34\bar2} & \vev{2\bar4\bar31} & \vev{3124} & \vev{42\bar1\bar3} \\
 \vev{1\bar3\bar4\bar2} & \vev{2\bar431} & \vev{31\bar24} & \vev{421\bar3} \\
 \vev{1\bar3\bar4\bar2} & \vev{24\bar31} & \vev{312\bar4} & \vev{4\bar213}
\end{array}
\end{equation}
and to obtain all the matrices associated with the $\{ TM \}$-set one simply
introduces a factor of $\pm$ in front of each matrix.

The complete list of L-matrices associated with the $\{VM_1 \}$-set begins by giving the
listing
\begin{equation}
\begin{array}{cccc}
 \vev{1432} & \vev{23\bar4\bar1} & \vev{3\bar2\bar14} & \vev{4\bar12\bar3} \\
 \vev{1432} & \vev{2\bar34\bar1} & \vev{32\bar1\bar4} & \vev{4\bar1\bar23} \\
 \vev{1\bar432} & \vev{234\bar1} & \vev{3\bar2\bar1\bar4} & \vev{41\bar23} \\
 \vev{1\bar432} & \vev{2\bar3\bar4\bar1} & \vev{32\bar14} & \vev{412\bar3} \\
 \vev{14\bar32} & \vev{234\bar1} & \vev{3\bar214} & \vev{4\bar1\bar2\bar3} \\
 \vev{14\bar32} & \vev{2\bar3\bar4\bar1} & \vev{321\bar4} & \vev{4\bar123} \\
 \vev{1\bar4\bar32} & \vev{23\bar4\bar1} & \vev{3\bar21\bar4} & \vev{4123} \\
 \vev{1\bar4\bar32} & \vev{2\bar34\bar1} & \vev{3214} & \vev{41\bar2\bar3} \\
\end{array}
 \qquad\qquad
\begin{array}{cccc}
 \vev{143\bar2} & \vev{23\bar41} & \vev{3\bar2\bar1\bar4} & \vev{4\bar123} \\
 \vev{143\bar2} & \vev{2\bar341} & \vev{32\bar14} & \vev{4\bar1\bar2\bar3} \\
 \vev{1\bar43\bar2} & \vev{2341} & \vev{3\bar2\bar14} & \vev{41\bar2\bar3} \\
 \vev{1\bar43\bar2} & \vev{2\bar3\bar41} & \vev{32\bar1\bar4} & \vev{4123} \\
 \vev{14\bar3\bar2} & \vev{2341} & \vev{3\bar21\bar4} & \vev{4\bar1\bar23} \\
 \vev{14\bar3\bar2} & \vev{2\bar3\bar41} & \vev{3214} & \vev{4\bar12\bar3} \\
 \vev{1\bar4\bar3\bar2} & \vev{23\bar41} & \vev{3\bar214} & \vev{412\bar3} \\
 \vev{1\bar4\bar3\bar2} & \vev{2\bar341} & \vev{321\bar4} & \vev{41\bar23}
\end{array}
\end{equation}
and to obtain all the matrices associated with the $\{ VM_1 \}$-set one simply
introduces a factor of $\pm$ in front of each matrix.

The complete list of L-matrices associated with the $\{VM_2 \}$-set begins by giving the
listing
\begin{equation}
\begin{array}{cccc}
 \vev{1243} & \vev{2\bar13\bar4} & \vev{34\bar2\bar1} & \vev{4\bar3\bar12} \\
 \vev{1243} & \vev{2\bar1\bar34} & \vev{3\bar42\bar1} & \vev{43\bar1\bar2} \\
 \vev{12\bar43} & \vev{2\bar134} & \vev{3\bar4\bar2\bar1} & \vev{431\bar2} \\
 \vev{12\bar43} & \vev{2\bar1\bar3\bar4} & \vev{342\bar1} & \vev{4\bar312} \\
 \vev{124\bar3} & \vev{2\bar134} & \vev{34\bar21} & \vev{4\bar3\bar1\bar2} \\
 \vev{124\bar3} & \vev{2\bar1\bar3\bar4} & \vev{3\bar421} & \vev{43\bar12} \\
 \vev{12\bar4\bar3} & \vev{2\bar13\bar4} & \vev{3\bar4\bar21} & \vev{4312} \\
 \vev{12\bar4\bar3} & \vev{2\bar1\bar34} & \vev{3421} & \vev{4\bar31\bar2} \\
\end{array}
 \qquad\qquad
\begin{array}{cccc}
 \vev{1\bar243} & \vev{213\bar4} & \vev{3\bar4\bar2\bar1} & \vev{43\bar12} \\
 \vev{1\bar243} & \vev{21\bar34} & \vev{342\bar1} & \vev{4\bar3\bar1\bar2} \\
 \vev{1\bar2\bar43} & \vev{2134} & \vev{34\bar2\bar1} & \vev{4\bar31\bar2} \\
 \vev{1\bar2\bar43} & \vev{21\bar3\bar4} & \vev{3\bar42\bar1} & \vev{4312} \\
 \vev{1\bar24\bar3} & \vev{2134} & \vev{3\bar4\bar21} & \vev{43\bar1\bar2} \\
 \vev{1\bar24\bar3} & \vev{21\bar3\bar4} & \vev{3421} & \vev{4\bar3\bar12} \\
 \vev{1\bar2\bar4\bar3} & \vev{213\bar4} & \vev{34\bar21} & \vev{4\bar312} \\
 \vev{1\bar2\bar4\bar3} & \vev{21\bar34} & \vev{3\bar421} & \vev{431\bar2}
\end{array}
\end{equation}
 and to obtain all the matrices associated with the $\{ VM_2 \}$-set one simply
introduces a factor of $\pm$ in front of each matrix.
 
The complete list of L-matrices associated with the $\{VM_3 \}$-set begins by giving the
listing
\begin{equation}
\begin{array}{cccc}
 \vev{1234} & \vev{2\bar1\bar43} & \vev{34\bar1\bar2} & \vev{4\bar32\bar1} \\
 \vev{1234} & \vev{2\bar14\bar3} & \vev{3\bar4\bar12} & \vev{43\bar2\bar1} \\
 \vev{123\bar4} & \vev{2\bar143} & \vev{3\bar4\bar1\bar2} & \vev{43\bar21} \\
 \vev{123\bar4} & \vev{2\bar1\bar4\bar3} & \vev{34\bar12} & \vev{4\bar321} \\
 \vev{12\bar34} & \vev{2\bar143} & \vev{341\bar2} & \vev{4\bar3\bar2\bar1} \\
 \vev{12\bar34} & \vev{2\bar1\bar4\bar3} & \vev{3\bar412} & \vev{432\bar1} \\
 \vev{12\bar3\bar4} & \vev{2\bar1\bar43} & \vev{3\bar41\bar2} & \vev{4321} \\
 \vev{12\bar3\bar4} & \vev{2\bar14\bar3} & \vev{3412} & \vev{4\bar3\bar21} \\
\end{array}
 \qquad\qquad
\begin{array}{cccc}
 \vev{1\bar234} & \vev{21\bar43} & \vev{3\bar4\bar1\bar2} & \vev{432\bar1} \\
 \vev{1\bar234} & \vev{214\bar3} & \vev{34\bar12} & \vev{4\bar3\bar2\bar1} \\
 \vev{1\bar23\bar4} & \vev{2143} & \vev{34\bar1\bar2} & \vev{4\bar3\bar21} \\
 \vev{1\bar23\bar4} & \vev{21\bar4\bar3} & \vev{3\bar4\bar12} & \vev{4321} \\
 \vev{1\bar2\bar34} & \vev{2143} & \vev{3\bar41\bar2} & \vev{43\bar2\bar1} \\
 \vev{1\bar2\bar34} & \vev{21\bar4\bar3} & \vev{3412} & \vev{4\bar32\bar1} \\
 \vev{1\bar2\bar3\bar4} & \vev{21\bar43} & \vev{341\bar2} & \vev{4\bar321} \\
 \vev{1\bar2\bar3\bar4} & \vev{214\bar3} & \vev{3\bar412} & \vev{43\bar21}
\end{array}
\end{equation}
and to obtain all the matrices associated with the $\{ VM_3 \}$-set one simply
introduces a factor of $\pm$ in front of each matrix.
$$~~$$

\end{document}

\bibliographystyle{elsart-numX}
\small\raggedright
\bibliography{Refs}

\end{document}